\documentclass[aps, pre, twocolumn, showpacs, showkeys]{revtex4}

\usepackage{graphicx}% Include figure files
\usepackage{bm}% bold math
\usepackage{amsmath}
\usepackage{amssymb}

\newcommand{\comment}[1]{}  % THIS DOESN'T SEEM TO BE ALLOWED!
\newcommand{\mtiny}[1] {{\mbox{\tiny #1}}} % THIS DOESN'T SEEM TO BE ALLOWED!
\newcommand{\mtn}[1] {{\mbox{\tiny #1}}} % THIS DOESN'T SEEM TO BE ALLOWED!
\newcommand{\myvec}[1]{\bf #1} % THIS DOESN'T SEEM TO BE ALLOWED!
\newcommand{\pp}{{\myvec{P}}}

\newcommand{\lav}{\left<}
\newcommand{\rav}{\right>}
\newcommand{\h}{\mathcal{H}}
\newcommand{\n}{Z}

\newcommand{\tn}{\tau_\mtiny{N}} % THIS DOESN'T SEEM TO BE ALLOWED!
\newcommand{\tB}{\tau_\mtiny{B}} % THIS DOESN'T SEEM TO BE ALLOWED!
\newcommand{\ta}{\tau_1} 
\newcommand{\tb}{\tau_2}
\newcommand{\tc}{\tau_3}
\newcommand{\td}{\tau_4}
\newcommand{\te}{\tau_5}
\newcommand{\tf}{\tau_6}

\begin{document}

\title{Analysis of the intraspinal calcium dynamics and its implications on the plasticity of spiking neurons}

\author{Luk C. Yeung}
\affiliation{Brown University Physics Department\\Box 1843 Providence, RI, 02912}

\author{Gastone C. Castellani}
\affiliation{Dip. Fisica Universit\`a Bologna \\Viale Berti Pichat 6/2, Bologna, Italy, 40127}

\author{Harel Z. Shouval}
\email{Harel_Shouval@brown.edu}
\affiliation{Brown University Physics Department\\ and Institute for Brain and Neural Systems\\Box 1843 Providence, RI, 02912}

\date{\today}

\begin{abstract}
The influx of calcium ions into the dendritic spines through the N-methyl-D-aspartate (NMDA) channels is believed to be the primary trigger for various forms of synaptic plasticity. In this paper, the authors calculate analytically the mean values of the calcium transients elicited by a spiking neuron undergoing a simple model of ionic currents and back-propagating action potentials. The relative variability of these transients, due to the stochastic nature of synaptic transmission, is further considered using a simple Markov model of NMDA receptors. One finds that both the mean value and the variability depend on the timing between pre- and postsynaptic action-potentials. These results could have implications on the expected form of synaptic-plasticity curve and can form a basis for a unified theory of spike time-dependent, and rate based plasticity.
\end{abstract}


\maketitle

%%%%%%%%
\section{Introduction}
%%%%%%%%

Calcium ions are ubiquitous mediators of metabolic change in many cellular systems. In the cortex, it is believed to be the primary chemical signal for the induction of bidirectional synaptic plasticity \cite{BearMalenka94, CummingsEtAl96, YangEtAl99}, which is widely accepted to be the basis for many forms of learning, memory and development. 

Experimentally, bi-directional synaptic plasticity can be induced by various different induction protocols \cite{DudekBear92, BlissCollingridge93,FeldmanEtAl98}, including the recently characterized spike-time-dependent plasticity (STDP) \cite{MarkramEtAl97,BiPoo98}. In this induction method, if a presynaptic action potential occurs within a window of a few tens milliseconds {\it before} a postsynaptic action potential ($\Delta t > 0$), long-term potentiation (LTP) is elicited; if the order is switched ($\Delta t < 0$), long-term depression (LTD) happens. Interestingly, STDP and many other induction protocols share a common property, which is their dependence on the amount of integrated activation of the N-methyl-D-aspartate receptors (NMDAR) during conditioning. NMDAR are natural coincidence detectors; their activation relies both on the efficiency of neurotransmitter release from the presynaptic cell, and on the level of depolarization of the postsynaptic cell. Most remarkably, they are permeable to calcium ions, which suggests a functional link between calcium influx through NMDAR and the induction of LTP and LTD. 

Recently, several dynamical models have been proposed to account for STDP \cite{SennEtAl01,KitajimaHara00,KarmarkarBuonomano02,AbarbanelEtAl02}, the relationship between STDP and rate-based models has been examined using averaging methods \cite{KempterEtAl99,ShouvalEtAl02b}, and there have been attempts to account for receptive field formation on the basis of STDP \cite{SongEtAl2000}. We have recently proposed a unified theory of synaptic plasticity \cite{ShouvalEtAl02} based on the following assumptions: (1) the level of intracellular calcium concentration determines the sign and magnitude of synaptic plasticity \cite{BearEtAl87, Lisman89,Artola-Singer-93}: when calcium falls below a threshold $\theta_d$, no plasticity occurs, when it falls between $\theta_d$ and $\theta_p$, $\theta_p > \theta_d$, LTD is induced, and for calcium above $\theta_p$, LTP is induced; (2) the relevant sources of calcium are the NMDA channels; and (3) dendritic back-propagating action potentials (BPAPs) contributing to STDP have a slow "after-depolarizing" tail component. We have shown that these simple assumptions are sufficient to account for the various experimental plasticity-induction protocols. In addition, this model has also produced previously uncharacterized predictions, such as: (a) the shape of the STDP learning curve should be frequency-dependent; and (b) there should be a novel form of spike time-dependent LTD for $\Delta t$ larger than the LTP-inducing intervals. Recent experimental results are consistent with these predictions \cite{NishiyamaEtAl00,SojstromEtAl01}.

In this paper, we present a systematic calculation of the calcium transients as a function of pre- and post-spike timings, as well as of the neuronal firing rates. In section \ref{sec:Ca_dyn} we derive the solution for the mean calcium transients, in the simplest case where the BPAP consists of a single exponential, and show that its dependence on $\Delta t$ is rather intuitive. A more realistic approach is taken in section \ref{sec:Ca_dyn2}, where the BPAP is composed of a sum of two exponentials with different time constants. This more complex assumption is in closer agreement with our previous work \cite{ShouvalEtAl02} and is consistent with experimental observations \cite{MageeJohnston97}. The analysis shows that the slow component contributes for different calcium levels in the baseline condition (pre only) and the post-pre condition ($\Delta t<0$), while the fast component sharpens the difference of the calcium levels in the transition  between post/pre ($\Delta t<0$) and pre/post ($\Delta t>0$) conditions. The sharpness of this transition is shown to be further enhanced by the appropriate choice of the calcium decay time-constant. For completeness, in section \ref{sec:rate}, we study the effects of calcium-accumulation due to repetitive firing which affects the rate-dependence of some plasticity-induction protocols including the STDP learning window. Finally, in section \ref{sec:var} we take into account the stochastic properties of synaptic transmission and estimate the trial-by-trial variability of the calcium transients. This variability can be significant if the number of postsynaptic NMDAR in each spine is small. Our analysis demonstrates that this variability depends $\Delta t$; increasing with $\Delta t$ for $\Delta t>0$. These results could significantly alter the form of the STDP curve under different experimental conditions.

%%%%%%%%
\section{Analysis of calcium influx through NMDAR\label{sec:Ca_dyn}}
%%%%%%%%

The postsynaptic intraspinal calcium ions are removed from these compartments both through diffusion and through binding to calcium buffers. Let these phenomena be characterized by a single time constant $\tau$, and assume that the calcium dynamics follow a first-order linear differential equation,

\begin{equation}
{\frac{d[Ca]}{dt}}=I(t)-\frac{1}{\tau}[Ca].
\label{eq:Ca}
\end{equation}
where $[Ca]$ is the intracellular calcium concentration, $I(t)$ is the inward calcium current, and $\tau$ is the time constant of the passive decay process. This simple dynamics is to the first approximation consistent with experiments \cite{MarkramEtAl95} \cite{SabatiniEtAl02}. Given a initial condition $[Ca](0)$, equation \ref{eq:Ca} has a solution of the form:

\begin{eqnarray}
[Ca](t)=e^{-t/\tau}\left[\int_0^te^{t/\tau}I(t^\prime)dt^\prime + [Ca](0)\right].
\label{eq:Ca_solution}\end{eqnarray}

Notice that if the calcium current is a superposition of separate current sources, the calcium concentration can also be decomposed into such a sum.

In this paper we consider the NMDA channels as the sole source of calcium currents. These channels are voltage dependent \cite{JahrStevens90}, and have a long open time that can usually be described as a sum of two exponentials \cite{CarmignotoVicini92}, thus its currents can be generally described as:

\begin{equation}
I(t,\{t^{pre}\},\{t^{post}\})= \bar{g} \h (V) f,
\label{eq:I}
\end{equation}
where $\{t^{pre}\}$ and $\{t^{post}\}$ are the sets of times of pre- and postsynaptic action-potentials (APs) that occured prior to time $t$ and $\bar{g}$ is the mean total conductance of the population of NMDAR in a synapse. The $\h$-function represents the voltage dependence of calcium influx through NMDAR, and $f$ is the fraction of NMDAR in open state. It is easy to see that $V = V(t, \{t^{pre} \}, \{ t^{post}\})$ and $f = f(t, \{t^{pre}\})$, where the time dependence of $f$ is a double exponential. 

In this section, however, we will focus on the simple scenario of a single pair of pre and postsynaptic APs, where the dynamics of the NMDAR and the BPAP are described as single exponentials with time constants $\tn$ and $\tB$, respectively. We will leave the more realistic two-component-BPAP case to the next section. Unless otherwise mentioned, we replace the stochastic variable $f$ by its mean without change of notation. Thus, for a single presynaptic spike at time $t^{pre}$, we can write,

\begin{equation}
f(t, t^{pre}) = \mu \Theta(t - t^{pre}) e^{-(t - t^{pre})/\tn},
\label{eq:f}
\end{equation}
where $\Theta$ is the Heaviside function and $\mu$ is the probability that a channel will open given a presynaptic AP. Let us assume that the postsynaptic depolarization is dominated by the BPAP. Therefore, $V$ in equation \ref{eq:I} can be written as:

\begin{align}
V(t, t^{post}) &= V_R + \mbox{BPAP(t)} \\
               &= V_R + V_B \Theta(t - t^{post})e^{-(t-t^{post})/\tB},
\label{eq:V}
\end{align}
where $V_R$ is resting membrane potential and $V_B$ is the magnitude of the BPAP due to a postsynaptic AP occuring at time $t^{post}$. In general, $\h$ is a non-linear and non-monotonic function of $V$. For tractability, we will make a major simplification by linearizing the $\h$-function in the interval $[V_R, V_B]$, 

\begin{equation}
\h(V) = a + bV. 
\label{eq:H}
\end{equation}

The original $\h$ function used and its linear fit are displayed in figure \ref{fig:comparison}. In appendix \ref{app:B}, we assess the validity of this approximation.
By using equations \ref{eq:f} through \ref{eq:H}, equation \ref{eq:I} becomes:

\begin{align}
I(t, t^{pre}, & t^{post}) \notag \\ 
= &  \bar{g} \left( a + b V_R + b V_B \Theta(t-t^{post})e^{-(t-t_{post})/\tB}\right) \notag \\
  &  \times \mu \Theta(t-t^{pre})e^{-(t-t^{pre})/\tn} \notag \\
= &  \bar{g} \mu \left(a + b V_R \right)\Theta(t-t^{pre})e^{-(t-t^{pre})/\tn} \notag\\
  &  + \bar{g} \mu b V_B \Theta(t-t^{post}) \Theta(t-t^{pre}) \notag \\
  &  \phantom{+ \mu} \times e^{-(t-t^{pre})/\tn} e^{-(t-t^{post})/\tB} \notag \\
= &  I^{pre}(t) + I^{+/-}(t).  \label{eq:I2}
\end{align}

The current separates into two components, one that depends only on the timing of the presynaptic AP ($I^{pre}$) and a second, associative term, that depends on the relative timing between the pre- and postsynaptic APs ($I^{+/-}$). The associative term will give rise to the differences between pre/post and post/pre conditions, which is essential to the STDP learning.

Let $\Delta t=t^{post}-t^{pre}$. It is straightforward to show that:

\begin{align}
I^{+/-} = \left\{ \begin{array}{l l}
I^+_{peak}\Theta(t-t^{post})e^{-(t-t^{post})/\ta}, & \mbox{if } \Delta t > 0 \\
\\
I^-_{peak}\Theta(t-t^{pre})e^{-(t-t^{pre})/\ta}, & \mbox{otherwise},
\end{array}\right.
\label{eq:Iprepost}
\end{align}
where $\ta^{-1} = \tB^{-1} + \tn^{-1}$, and $I^+_{peak} = I^+_{peak}(\Delta t)$ and $I^- = I^-(\Delta t)$ denote the peak magnitudes of the pre/post component of the calcium current for $\Delta t>0$ and $\Delta t<0$ respectively:

\begin{subequations}
\begin{align}
I^+_{peak}(\Delta t) & = \bar{g}\mu b V_B e^{-\Delta t/\tau_\mtn{N}} \label{eq:I+}\\
I^-_{peak}(\Delta t) & = \bar{g}\mu b V_B e^{\Delta t/\tau_\mtn{B}}. \label{eq:I-}
\end{align}
\end{subequations}

\begin{figure}
\includegraphics[width=7cm]{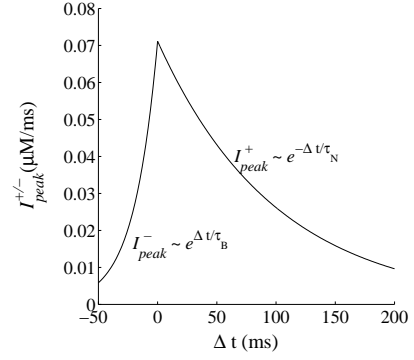}
\caption{\label{fig:Ipeak} Associative component of the peak calcium current as a function of $\Delta t$. $\bar{g} b= 1.5 \times 10^{-3}$, $\mu = 0.8$, $V_B = 60$ mV, $\tn = 100$ ms and $\tB = 20$ ms.}
\end{figure}

The dependence of the peak current on $\Delta t$ is quite simple (FIG. \ref{fig:Ipeak}). For $\Delta t>0$ it decays exponentially with the time constant of the NMDAR open probability $\tn$; whereas for $\Delta t<0$ it decays with $\tB$, the width of the BPAP. It is straightforward to see that, if $\Delta t < 0$ yields calcium levels corresponding to LTD, and if to each calcium concentration is associated a degree of plasticity, there should be a $\Delta t > 0$ region that also corresponds to LTD. The difference is that, since $\tB << \tn$, $\theta_d$ in the post/pre region is reached for values of $\Delta t$ much closer to zero than it is for the pre/post region. Such difference will be even greater in the case where the BPAP's dynamic is described by two time constants instead of one (see section \ref{sec:Ca_dyn2}).

Using equations \ref{eq:I2} and \ref{eq:Iprepost}, we can rewrite equation \ref{eq:Ca_solution} in terms of each of the components of the calcium current,

\begin{multline}
[Ca](t) = [Ca]^{pre}(t)+ \Theta(\Delta t)[Ca]^+(t)\\
        + \Theta(-\Delta t)[Ca]^-(t)+ [Ca](0) e^{-t/\tau}.
\label{eq:Ca_av1}
\end{multline}
with

\begin{subequations}
\begin{align}
&[Ca]^{pre}(t) = e^{-t/\tau}\int I^{pre}(t') e^{t'/\tau} dt' \label{eq:CaPre}\\
&[Ca]^+(t) = e^{-t/\tau}\int I^+(t') e^{t'/\tau} dt' \label{eq:Ca+}\\
&[Ca]^-(t) = e^{-t/\tau}\int I^-(t') e^{t'/\tau} dt' \label{eq:Ca-}.
\end{align}
\end{subequations}

Integrating explicitly, we have,

\begin{align}
&[Ca]^{pre}(t) = \bar{g} \mu (a + b V_R) \tb \Theta(t-t^{pre}) \tag{\ref{eq:CaPre}}\\
&\phantom{[Ca]^{pre}(t) = } \times \left( e^{-(t-t^{pre})/\tn}-e^{-(t-t^{pre})/\tau}\right) \notag \\
&[Ca]^+(t) = I^+_{peak} \tc \Theta(t-t^{post}) \tag{\ref{eq:Ca+}} \\
&\phantom{[Ca]^{pre}(t) = } \times \left(e^{-(t-t^{post})/\ta}-e^{-(t-t^{post})/\tau}\right) \notag\\
&[Ca]^-(t) = I^-_{peak} \tc \Theta(t-t^{pre}) \tag{\ref{eq:Ca-}} \\
&\phantom{[Ca]^{pre}(t) = } \times \left(e^{-(t-t^{pre})/\ta}-e^{-(t-t^{pre})/\tau}\right) \notag,
\end{align}
where $\tb^{-1}=\tau^{-1} - \tn^{-1}$ and $\tc^{-1} = \tau^{-1}-\tn^{-1}-\tB^{-1}$.

\begin{figure*}
\includegraphics[width=7cm]{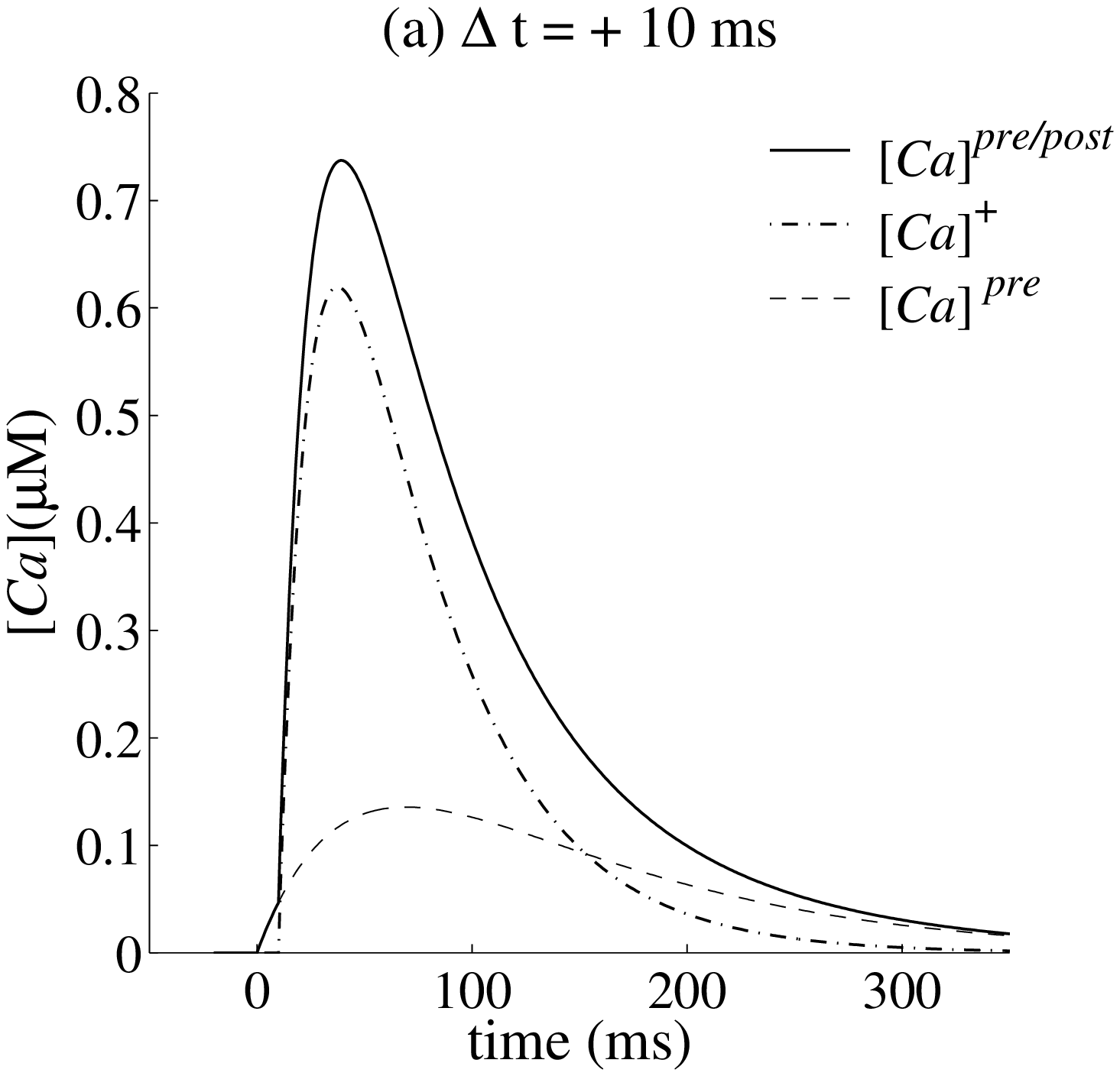}
\hspace{0.5cm}
\includegraphics[width=7cm]{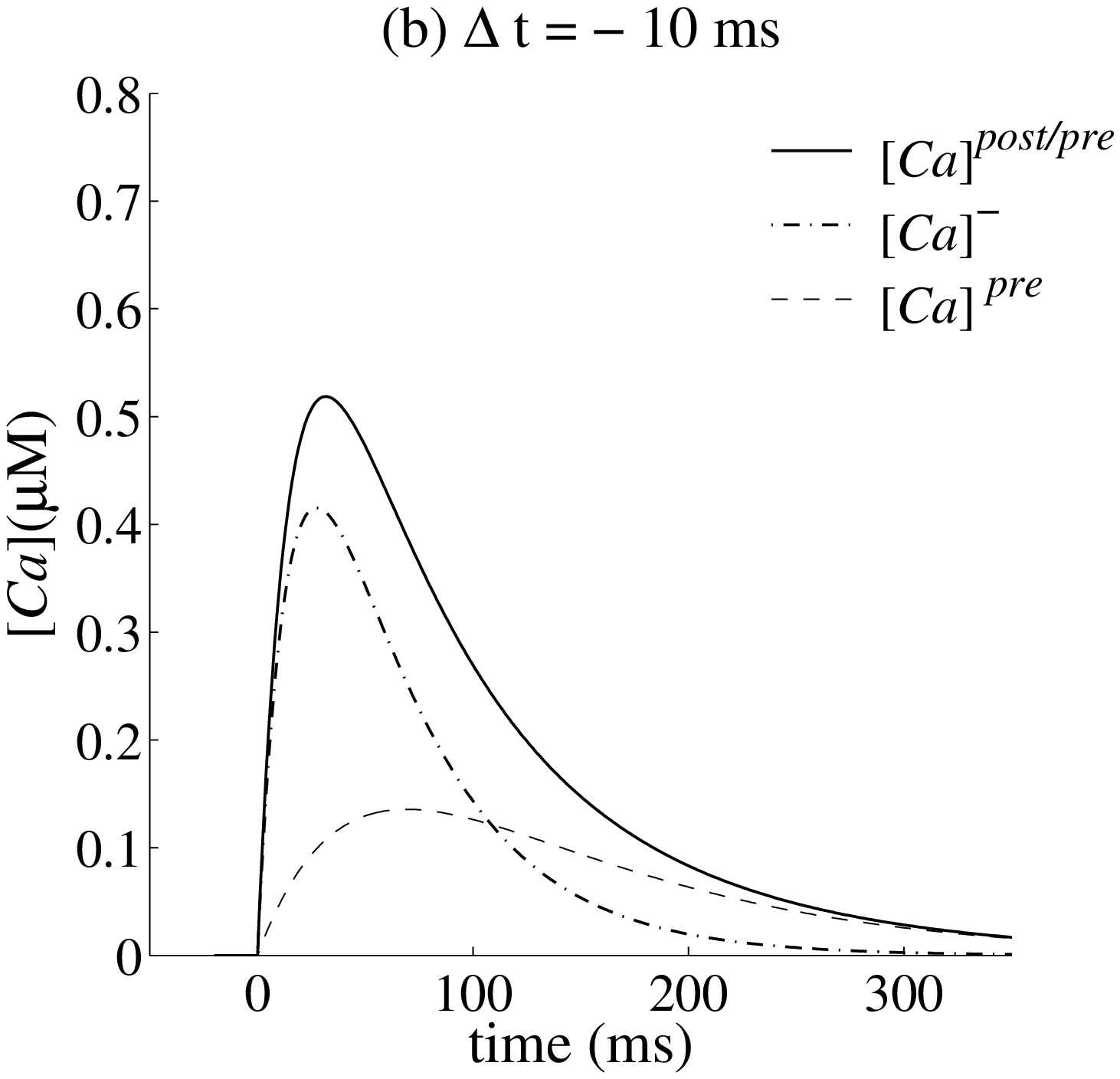}
\caption{\label{fig:Ca_dyn}The dynamics of the total calcium (solid line) is composed of two different contributions: the calcium due to the pre-spike alone (dashed line) and the calcium due to the association of pre- and postspikes (dash-dot line). $a\bar{g} = 1.03 x 10^{-1}$, $V_R = -65 mV$, $\tau=50$ ms, $[Ca](0) = 0$ and the onset of $t_{pre}$ is at time zero. The remaining parameters are as in FIG. \ref{fig:Ipeak}. a) $\Delta t=10$ ms, b) $\Delta t=-10$ ms. 
}
\end{figure*}

Examples of calcium transients are displayed in FIG. \ref{fig:Ca_dyn}, where $[Ca]^{pre/post} = [Ca]^{pre} + [Ca]^+$ and $[Ca]^{post/pre} = [Ca]^{pre} + [Ca]^-$ . Naturally, the $[Ca]^{pre}$ term is identical for the pre-post (FIG. \ref{fig:Ca_dyn}a, $\Delta t$ = 10 ms) and the post-pre (FIG. \ref{fig:Ca_dyn}b, $\Delta t$ = -10 ms) conditions (dashed line). The difference between these conditions lies in the associative term (dash-dot line). $[Ca]^+$ raises at the point $t = t^{post}$ and $[Ca]^-$ raises at $t = t^{pre}$, their relative amplitudes being determined by $I^{+/-}_{peak}$ (FIG. \ref{fig:Ipeak}). 

%%%%%%%%
\section{A back-propagating action potential with two components \label{sec:Ca_dyn2}}
%%%%%%%%

\begin{figure*}
\includegraphics[width=7cm]{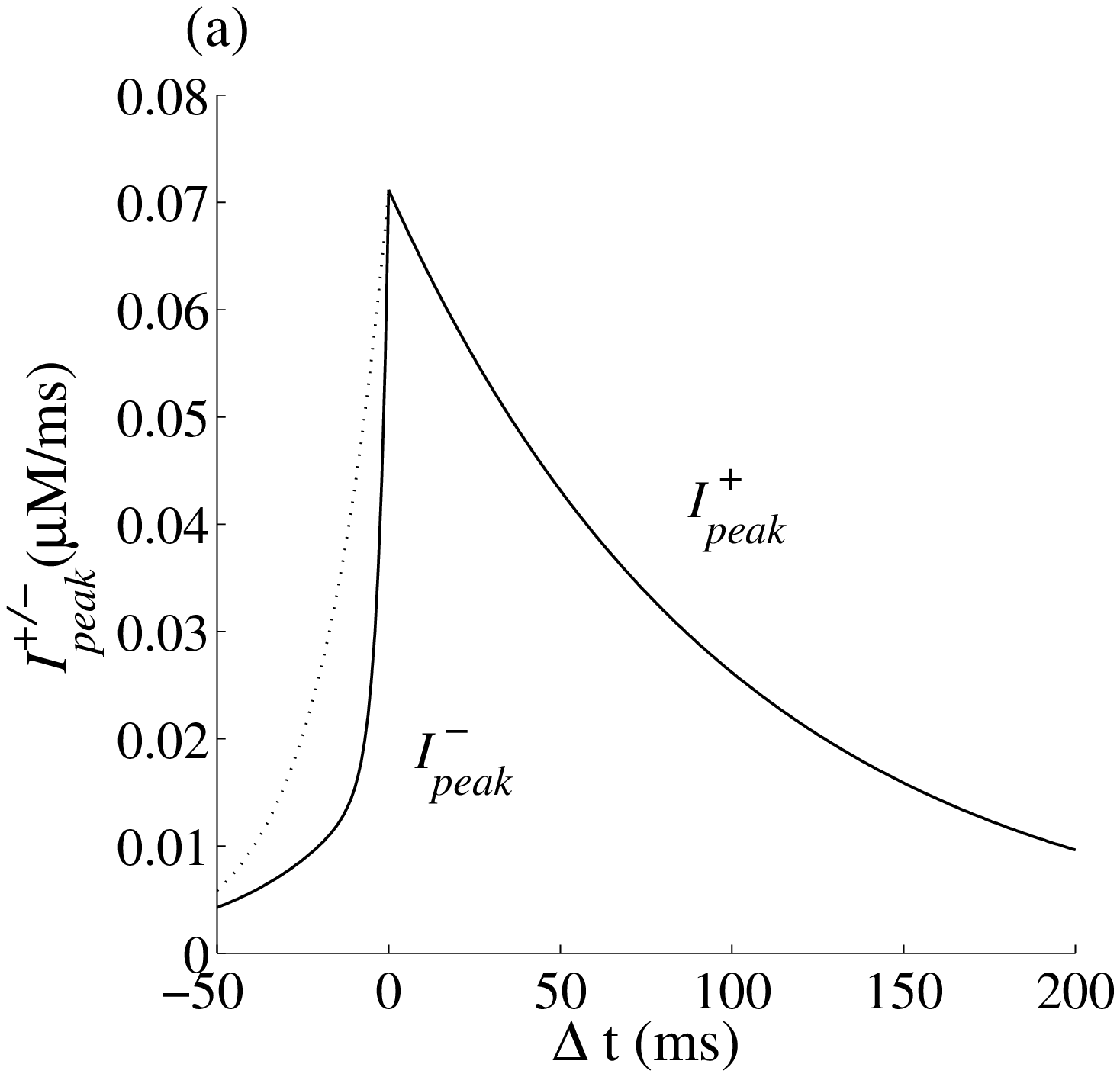}
\hspace{0.5cm}
\includegraphics[width=7cm]{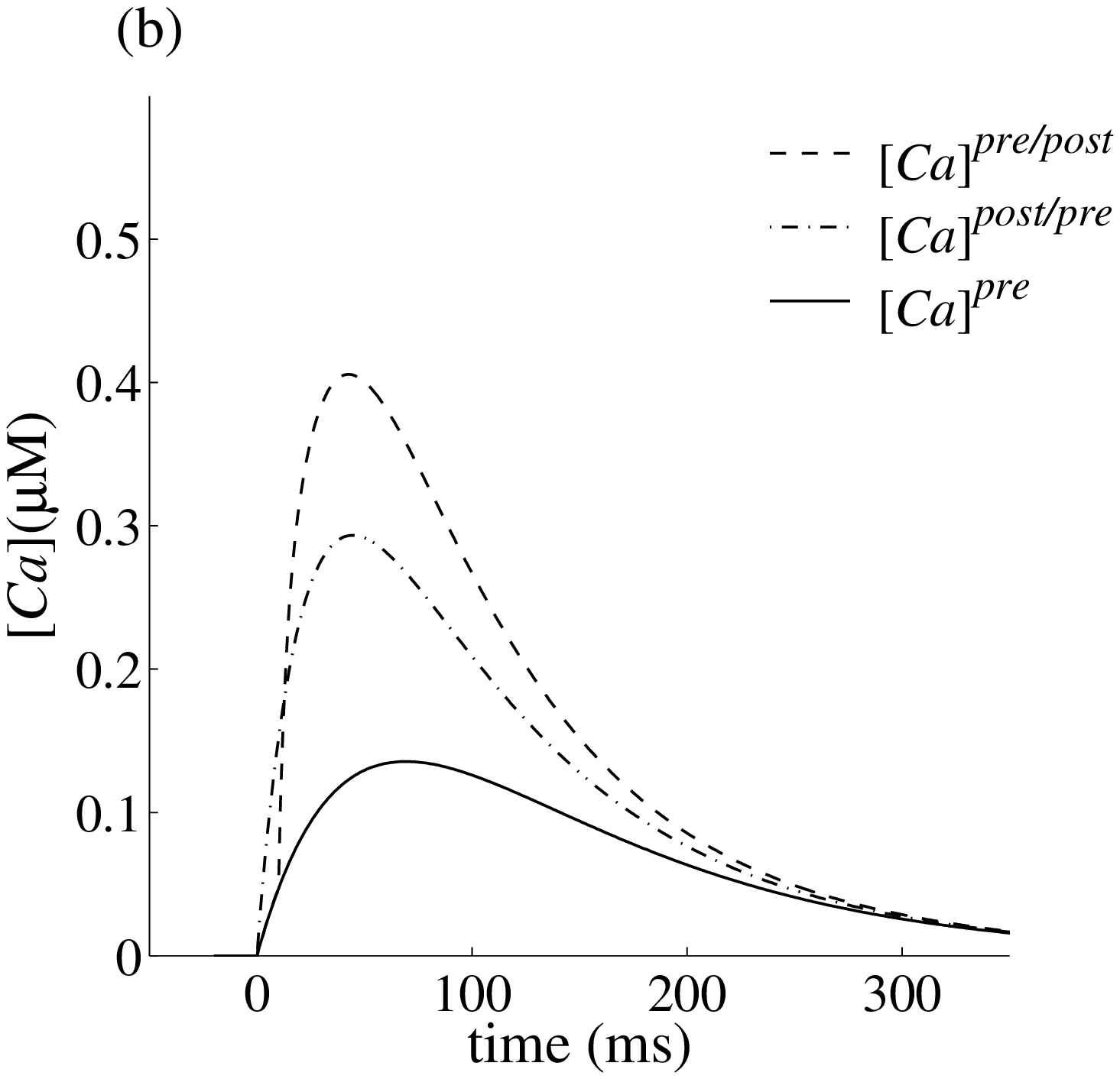}
\caption{\label{fig:Ipeak_3}
a) Associative component of the peak calcium current as a function of $\Delta t$, for $v^f = 0.75$, $v^s = 0.25$, $\tB^f = 3$ ms and $\tB^s = 35$ ms. The dotted line shows the previous, single BPAP component result, for comparison. b) Total calcium transients for $\Delta t = 10$ ms (dashed line), -10 ms (dashed-dot line) and $[Ca]^{pre}$ (solid line). All the other parameters are as in FIG. \ref{fig:Ipeak}, \ref{fig:Ca_dyn}.
}
\end{figure*}

The model described in the previous section assumes a BPAP with a single exponential falling phase. However, there are experimental indications that an additional slow after-depolarizing component exists in some cell bodies \cite{Feldman00} and dendrites \cite{MageeJohnston97,LarkumEtAl01}. The dynamics of the BPAP defines to what extent the postsynaptic spike-timing interacts with the presynaptic one; in particular, a slower decaying process ensures that such interaction spans for a period of tens of milliseconds, as required by the STDP learning rule. In this section, we assume a BPAP composed by a sum of two exponentials, with, respectively, a slow $\tB^s$ and a fast $\tB^f$ time constants. Equation \ref{eq:V} should therefore be re-written as:

\begin{align} 
V(t, t^{post}) = & V_R + V_B \Theta(t-t^{post}) \tag{\ref{eq:V}'}\\
&\times \left[ v^f e^{-(t-t^{post})/\tB^f } + v^s e^{-(t-t^{post})/ \tB^s } \right] \notag
\end{align}
where $v^s$, $v^f$, $\tB^s$ and $\tB^f$ are the relative amplitudes and time constants of the slow and the fast components of the BPAP respectively, $(v^s, v^f)>0, v^s+v^f=1$. Analogously to the previous section, the calcium currents from equation \ref{eq:Iprepost} can be separated into two components, $I^{pre}$ and $I^{+/-}$. $I^{pre}$ is identical to what we have derived before, whereas $I^{+/-}$ reads:

\begin{align}
I^{+/-}(t) = \left\{\begin{array}{l r} 
\Theta(t-t^{post})I^{+}_{peak}\left[  v^f e^{-(t-t^{post})/\ta^f} \right. \\
\phantom{\Theta(t)} \left.+ v^s e^{-(t-t^{post})/\ta^s} \right], \phantom{blah} \mbox{if } \Delta t > 0 \\
\\\tag{\ref{eq:Iprepost}'}
\Theta(t-t^{pre})\left[  I^{-f}_{peak} v^f e^{-(t-t^{pre})/\ta^f} \right. \\
\phantom{\Theta(t)} \left.+ I^{-s}_{peak} v^s e^{-(t-t^{pre})/\ta^s} \right], \phantom{bl} \mbox{otherwise }  
\end{array}\right. 
\end{align}
where $I^{+}_{peak}$ is as defined in equation \ref{eq:I+}, $I^{-f}_{peak}=\bar{g} \mu b V_B e^{\Delta t/\tB^f}$, $I^{-s}_{peak}=\bar{g} \mu b V_B e^{\Delta t/\tB^s}.$ The calcium current $I^{pre/post}$ peaks at $t=t^{post}$ for $\Delta t>0$ with amplitude $I^+_{peak}$, and at $t=t^{pre}$ for $\Delta t<0$ with amplitude $I^-_{peak}=v^sI^{-s}_{peak} +v^f I^{-f}_{peak}$. Thus, $I^-_{peak}$ decays more abruptly than its single-component-BPAP counterpart (FIG. \ref{fig:Ipeak_3}a). 

We can again integrate each component of the calcium currents separately. We find for this case:

\begin{align}
&[Ca]^{pre}(t) = \bar{g} \mu(a+b V_R) \tb \Theta(t-t^{pre}) \tag{\ref{eq:CaPre}'} \\
   &\phantom{[Ca]^{pre}(t) = } 
\times \left( e^{-(t-t^{pre})/\tn}-e^{-(t-t^{pre})/\tau} \right) \notag \\
&[Ca]^+(t) = I^+_{peak} \Theta(t-t^{post}) \tag{\ref{eq:Ca+}'} \\
   &\phantom{[Ca]^{pre}(t) = } 
\times \left\{ v^f \tc^f \left( e^{-(t-t^{post})/\ta^f}-e^{-(t-t^{post})/\tau} \right) \right. \notag \\
   &\phantom{[Ca]^{pre}(t) = \times} 
       \left.  + v^s \tc^s \left( e^{-(t-t^{post})/\ta^s}-e^{-(t-t^{post})/\tau} \right) \right\} \notag \\
&[Ca]^-(t) = \Theta(t-t^{pre}) \tag{\ref{eq:Ca-}'} \\
   &\phantom{[Ca]^{pre}(t) = } 
\times \left\{ I^{-f}_{peak} v^f \tc^f \left( e^{-(t-t^{pre})/\ta^f}-e^{-(t-t^{post})/\tau} \right) \right. \notag\\
   &\phantom{[Ca]^{pre}_{peak}(t) = \times} 
       \left. + I^{-s} v^s \tc^s \left( e^{-(t-t^{pre})/\ta^s}-e^{-(t-t^{pre})/\tau} \right) \right\} \notag
\end{align}
where $(\ta^f)^{-1}=(\tB^f)^{-1}+\tn^{-1}$, $(\tc^f)^{-1}=\tau^{-1}-\tn^{-1}-(\tB^f)^{-1}$, and $\ta^s$ and $\tc^s$ are defined similarly to $\tB^s$, but with $\tB^s$ replacing $\tB^f$.

\begin{figure*}
\includegraphics[width=7cm]{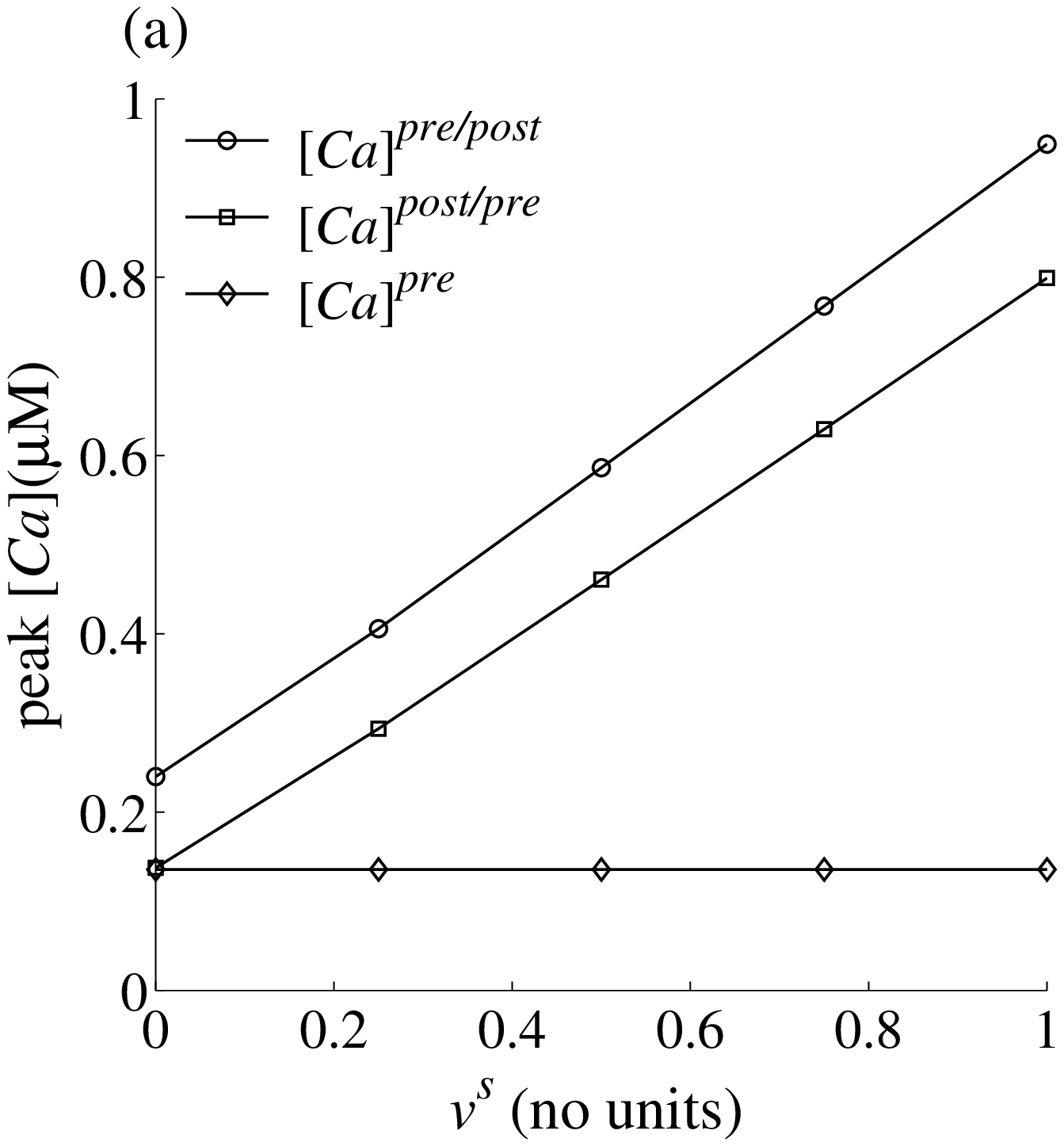}
\hspace{0.5cm}
\includegraphics[width=7cm]{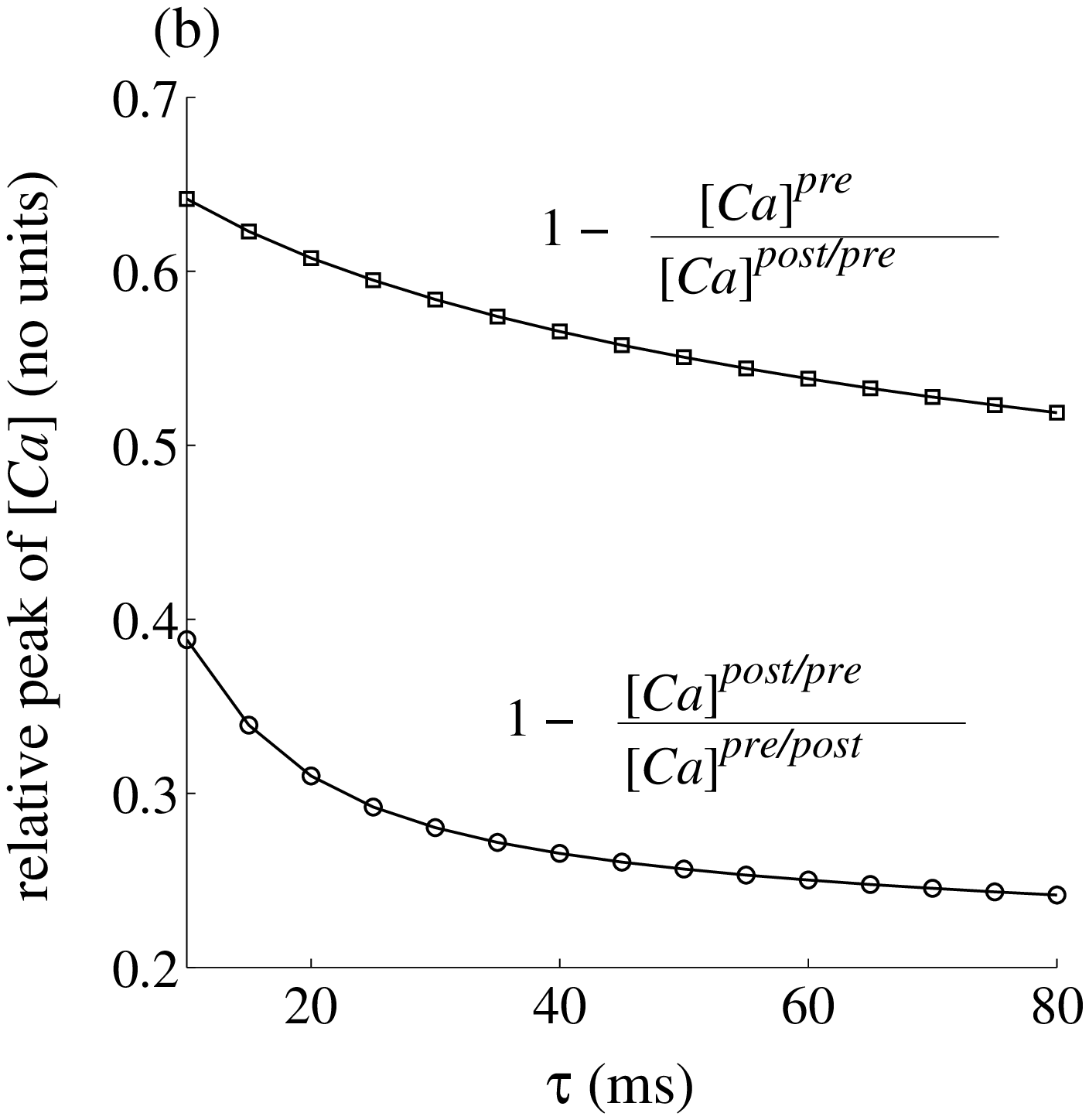}
\caption{\label{fig:Capeak_3}
a) Peak calcium transients as a function of the slow component of the BPAP $v^s$ for the pre-before-post (circle), post-before-pre (square) and the pre-only (diamond) conditions. b) Relative magnitudes of the peak calcium transients, between pre and post-pre (circle), and post-pre and pre-post (square) conditions, as a function of the calcium time-constant $\tau$. All parameters are as in FIG. \ref{fig:Ipeak_3}.}
\end{figure*}

The effects of the BPAP slow component on the calcium transients is shown in FIG. \ref{fig:Ipeak_3}b and \ref{fig:Capeak_3}a. Since both $[Ca]^+$ and $[Ca]^-$ have the same linear dependence on the magnitude $v_s$ of this slow component, it primarily contributes to the separation between the pre-only condition (diamond), and the associative conditions (circle and square). 

A longer decaying tail for the postsynaptic variable provides a longer range of time-interval in which interaction with the presynaptic variables is possible. Thus, the dynamics of the presynaptic variable, such as the calcium time constant $\tau$, will also influence the relative magnitudes of the calcium transients for the pre-post and post-pre conditions. Indeed, the combination of a faster calcium kinetics with the different BPAP time courses enhances the difference between these magnitudes (FIG. \ref{fig:Capeak_3}b). Therefore, a two-component-BPAP model produces a sharper transition of the learning curve between the post-pre LTD and the pre-post LTP. Initial estimates of the calcium time constant in the spines were of the order of 70-125ms \cite{MarkramEtAl95,MurthyEtAl00}. However, these values are obtained through calcium imaging, whose results may depend on the kinetics of the calcium indicators used, limiting the accuracy of the estimates. Recent experimental techniques, in which the effect of the calcium buffers are minimized, have lead to time constants of as low as 20 ms \cite{SabatiniEtAl02}.

%%%%%%%%
\section{Rate-dependence of the calcium transients \label{sec:rate}}
%%%%%%%%

In the previous sections we have analyzed the calcium transients evoked by one pair of pre and postsynaptic spikes, these have significant implications for STDP at low frequency. However, STDP at low frequency is not the sole method for inducing synaptic plasticity, and in some cases even STDP requires that the pairs of pre and postspikes are delivered above a certain frequency \cite{MarkramEtAl97,SojstromEtAl01}. Thus it is important to analyze the contribution of spike delivery frequency $\nu$, to the calcium concentration, in addition to the relative timing effect. We analyze the case in which both pre and postsynaptic neurons fire regularly at the same frequency $\nu$, the onset of the postsynaptic spike-train being shifted from the presynaptic spike train by $\Delta t < 1/\nu$. Notice, however, that in the most general case the number of presynaptic and postsynaptic spikes as well as their timing differences will vary.

Recall that the NMDA current in equation \ref{eq:I} is the product between the fraction of opened channels $f(t)$ and the linearized Mg-block function $\h(V(t)) = a + b V(t)$, where $V(t) = V_R + \mbox{BPAP}(t)$ omits depolarization due to the excitatory postsynaptic potentials (EPSPs). For a sequence $\{t^{pre}_1, t^{pre}_2, ..., t^{pre}_N\}$ of presynaptic spike-times and $\{t^{post}_1, t^{post}_2, ..., t^{post}_M\}$ of postsynaptic spike-times, we can write:

\begin{equation}
f(t) = \sum_{n=1}^N A_n \exp\left(-\frac{t-t_n^{pre}}{\tn}\right)\Theta \left( t-t_n^{pre} \right),
\label{eq:deff}
\end{equation}

\begin{align}
\mbox{BPAP} (t) =  V_B& \sum_{m=1}^{M-1}\left[ \exp \left(-\frac{t - t_m^{post}}{\tB}\right) \Theta \left(t - t_m^{post}\right) \notag \right.\\
 & \phantom{\sum_{m=1}^{M-1}\left[ \right]} \bigg. \times \Theta \left( -t + t_{m+1}^{post} \right)\bigg] \label{eq:BPAP}\\
 & + \exp \left( -\frac{t-t_M^{post}}{\tB}\right)\Theta \left( t - t_M^{post}\right), \notag
\end{align}
where $N$ and $M$ are the total number of pre and post-spikes before time $t$, and $A_n$ is the increase of the fraction of opened channels upon each pre-spike $n$. Let $f(t)$ be at level $B_n$ immediately before the pre-spike $n$ occurs. If the NMDA opening is proportional to the amount of non-opened channels, then $A_n = \mu (1-B_n)$, where $\mu$ is the fraction of previously closed channels that open due to the presynaptic spike, $\mu < 1$. Writing $B_n$ explicitly for each $n$, it is easy to see that it satisfies the following expression:

\begin{align}
B_n  =& \mu \sum_{i=1}^{n-1} e^{-i/\nu \tn} \notag \\
     =& \mu e^{-1/\nu \tn} \frac{1 - (1-\mu)^n e^{-n/\nu\tn}}{1 - (1-\mu) e^{-1/\nu\tn}} 
\end{align}
and

\begin{equation}
B_\infty= \lim_{n \to \infty}B_n= \frac{\mu e^{-1/\nu\tn}}{1 - (1-\mu)e^{-1/\nu\tn}}
\end{equation}

With this, equation \ref{eq:deff} can be written as:

\begin{align*}
f(t) = &\sum_{n=1}^{N-1} \left( B_n + A_n \right)\exp{\left(-\frac{t-t^{pre}_n}{\tn}\right)} \\ 
& \times \Theta(t - t^{pre}_n) \Theta(-t+t^{pre}_{n+1}) \\ 
+ & \left( A_N + B_N \right) \exp{\left( -\frac{t-t^{pre}_N}{\tn}\right)} \Theta (t - t^{pre}_N)
\end{align*}

\begin{align}
\phantom{----} = & \sum_n^{N-1} \left[ B_n(1-\mu) + \mu \right] \exp{\left(-\frac{t-t^{pre}_n}{\tn}\right)} \notag \\ 
& \times \Theta(t - t^{pre}_n) \Theta(-t+t^{pre}_{n+1}) \label{eq:final_f} \\
+ & \left[B_N (1 - \mu) + \mu \right] \exp{\left( -\frac{t-t^{pre}_N}{\tn}\right)} \Theta (t - t^{pre}_N). \notag
\end{align}

We can define, analogously to equation \ref{eq:I2} and the expressions \ref{eq:CaPre} through \ref{eq:Ca-} from section \ref{sec:Ca_dyn}, the following variables: $I^{pre} = \bar{g}(a+bV_R) f(t)$, $I^{+/-} = \bar{g} b \mbox{BPAP}(t) f(t)$ and:

\begin{subequations}
\begin{align}
[Ca]^{pre}(t) & = e^{-t/\tau} \int I^{pre}(t')e^{t'/\tau} dt' \label{eq:CaPre_multi},\\
[Ca]^{+/-}(t) & = e^{-t/\tau} \int I^{pre/post}(t') e^{t'/\tau} dt'. \label{eq:CaPrePost_multi}
\end{align}
\end{subequations}

Using equation \ref{eq:final_f}, equation \ref{eq:CaPre_multi} reads:

\begin{multline}
[Ca]^{pre} = \bar{g} e^{-t/\tau} \left( a + bV_R\right)\tb \\
  \times \left\{ \left(e^{1/\nu \tb}-1 \right) \sum_{n=1}^{N-1}\left[ B_n(1-\mu)+\mu\right]e^{t^{pre}_n/\tau} \right. \\
  + \left( e^{t/\tb} - e^{t^{pre}_N/\tb} \right) \left[ B_N(1-\mu) +\mu \right]  \\
  \Bigg. \times e^{t^{pre}_N/\tn} \Bigg\} .
\end{multline}

We use the approximation $B_n = B_\infty$, and write the time indexes with respect to the timing of the first spike:

\begin{subequations}
\begin{align}
& t^{pre}_n = t^{pre}_1 + \frac{n-1}{\nu} \label{eq:defa}, \\
& t^{pre}_N = t^{pre}_1 + \frac{\lfloor \nu t \rfloor - 1}{\nu} \mbox{ and } \label{eq:defb}\\
& N = \lfloor (t - t^{pre}_1) \nu \rfloor + 1, \label{eq:defc}
\end{align}
\end{subequations}
where $\lfloor~\rfloor$, is the floor operation. Evaluating the sum, we have:

\begin{align}
[Ca]^{pre} = & \bar{g} e^{-t/\tau}(a + bV_R)\tb\left[ B_{\infty}(1-\mu)+\mu\right] \notag \\ 
           & \times \left\{ \left( e^{1/(\nu \tb)}-1\right) e^{t^{pre}_1/\tau} \left[  \frac{1-\exp\left(\frac{N-1}{\nu \tau}\right)}{1 - \exp \left( \frac{1}{\nu \tau} \right)}\right] \right. \notag \\
           & \phantom{\left\{ \right\}} \Bigg. + \exp \left( \frac{t^{pre}_N}{\tn} + \frac{t}{\tb} \right) - \exp \left( \frac{t^{pre}_N}{\tau}\right) \Bigg\} ,
\end{align}
where $N$ and $t^{pre}_N$ are defined as in equations \ref{eq:defb} and \ref{eq:defc}.

To calculate the contribution to the calcium transients due to the interaction of the pre and postsynaptic spikes, we substitute the expression \ref{eq:BPAP} and \ref{eq:final_f} into equation \ref{eq:CaPrePost_multi}. Let $t^{post}_m \equiv t^{pre}_n+\Delta t$ and $t^{post}_M \equiv t^{pre}_N+\Delta t$. The indexes $m$ and $M$ will always refer to the post-spikes while $n$ and $N$ will always refer to the pre-spikes; for simplicity, we drop the superscripts so that $t_n^{pre} = t_n$ and $t_m^{post} = t_m$. Because the depolarization due to each post-spike does not build up, the product between the sums given by expressions \ref{eq:final_f} and \ref{eq:BPAP} will yield only four terms, due to the interaction of the pairs $\left\{ (n, m-1); (n, m); (N, M-1); (N, M) \right\}$. These terms are:

\begin{subequations}
\begin{align}
&\int_0^t f(t') \mbox{BPAP}(t') dt' = V_B\left[ B_\infty(1-\mu)+\mu \right] \Bigg\{ \Bigg. \notag \\ 
&\phantom{+} \sum_{n=2}^{N-1} \int_{t_n}^{t_m} \exp \left(-\frac{t'-t_n}{\tn} - \frac{t'-t_{m-1}}{\tB} + \frac{t'}{\tau} \right)dt' \label{eq:termd}\\
&+ \sum_{n=1}^{N-1} \int_{t_m}^{t_{n+1}} \exp \left(-\frac{t' - t_n}{\tn} - \frac{t'-t_m}{\tB} + \frac{t'}{\tau} \right) dt' \label{eq:terme}\\
&+ \int_{t_N}^{t_*} \exp \left(-\frac{t'-t_N}{\tn} - \frac{t-t_{M-1}}{\tB} + \frac{t'}{\tau} \right) dt' \label{eq:termf}\\
&+ \delta_{M, N} \int_{t_M}^{t} \exp \left(-\frac{t' - t_N}{\tn} - \frac{t' - t_M}{\tB} + \frac{t'}{\tau}\right) dt' \Bigg. \Bigg\} \label{eq:termg},
\end{align}
\end{subequations}
where $t_* = \min (t, t_M)$ and we have used that $B_n = B_N = B_\infty$. Recalling expressions \ref{eq:defa} through \ref{eq:defc}, one can easily perform the integrations \ref{eq:termd} through \ref{eq:termg}. The remaining sums are finite power series, therefore they converge. Thus, the calcium concentration due to the interaction between pre and postsynaptic spikes will be:

\begin{multline}
[Ca]^{+/-} = \bar{g} e^{- t/\tau} b V_B\left[ B_\infty(1-\mu)+\mu \right]\\
 \left\{ \mbox{\ref{eq:termd} + \ref{eq:terme} + \ref{eq:termf} + \ref{eq:termg}}\right\}
\end{multline}
with each of the terms being:

\begin{multline}
\mbox{\ref{eq:termd}} = \tc \left[ \exp \left(\frac{\Delta t}{\tb} - \frac{1}{\nu \tB}\right) - \exp \left( \frac{\Delta t - 1/\nu}{\tB}\right) \right] \notag \\
 \times e^{t_1^{pre}/\tau} \left[ \frac{1 - \exp \left( \frac{N-1}{\nu \tau}\right) }{1-\exp \left( \frac{1}{\nu \tau}\right) } \right] \notag
\end{multline}

\begin{multline}
\mbox{\ref{eq:terme}} = \tc \left[ \exp \left(\frac{\Delta t}{\tB} + \frac{1}{\nu \tc}\right) - \exp \left( \frac{\Delta t}{\tb}\right)\right] \notag \\
 \times e^{t_1^{pre}/\tau} \left[ \frac{1 - \exp \left( \frac{N-1}{\nu \tau}\right) }{1-\exp \left( \frac{1}{\nu \tau}\right) } \right] \notag
\end{multline}

\begin{align*}
\mbox{\ref{eq:termf}} = \left\{ \begin{array}{lc}
\tc \left[ \exp \left(\frac{\Delta t}{\tb} - \frac{1}{\nu \tB}\right) - \exp \left(\frac{\Delta t - 1/\nu}{\tB}\right) \right] \\
\phantom{-} \times  \exp \left(\frac{t_N^{pre}}{\tau} \right), \phantom{----|} \text{ if $\left( t^{pre}_N + \Delta t \right) < t$} \\
\\
\tc \left[ \exp \left(\frac{t^{pre}_N}{\ta} + \frac{t}{\tc}\right) - \exp \left(\frac{t^{pre}_N}{\tau}\right) \right] \\
\phantom{-} \times \exp \left( \frac{\Delta t - 1/\nu}{\tB}\right), \phantom{---|} \text{if $ \left( t^{pre}_N + \Delta t \right) > t$.} 
\end{array} \right.
\end{align*}

\begin{align*}
\mbox{\ref{eq:termg}} = \left\{ \begin{array}{lc} 
\tc \left[ \exp \left(\frac{t^{post}_M}{\ta} +\frac{t}{\tc}\right) - \exp \left(\frac{t_M^{post}}{\tau}\right) \right] \\
\phantom{-} \times  \exp \left(-\frac{\Delta t}{\tn} \right), \phantom{\text blah blah} \text{ if $\left( t^{pre}_N + \Delta t \right) < t$}  \\
\\
0 , \phantom{-----------} \text{if $ \left( t^{pre}_N + \Delta t \right) > t$.} 
\end{array} \right.
\end{align*}

\begin{figure*}
\includegraphics[width=7cm]{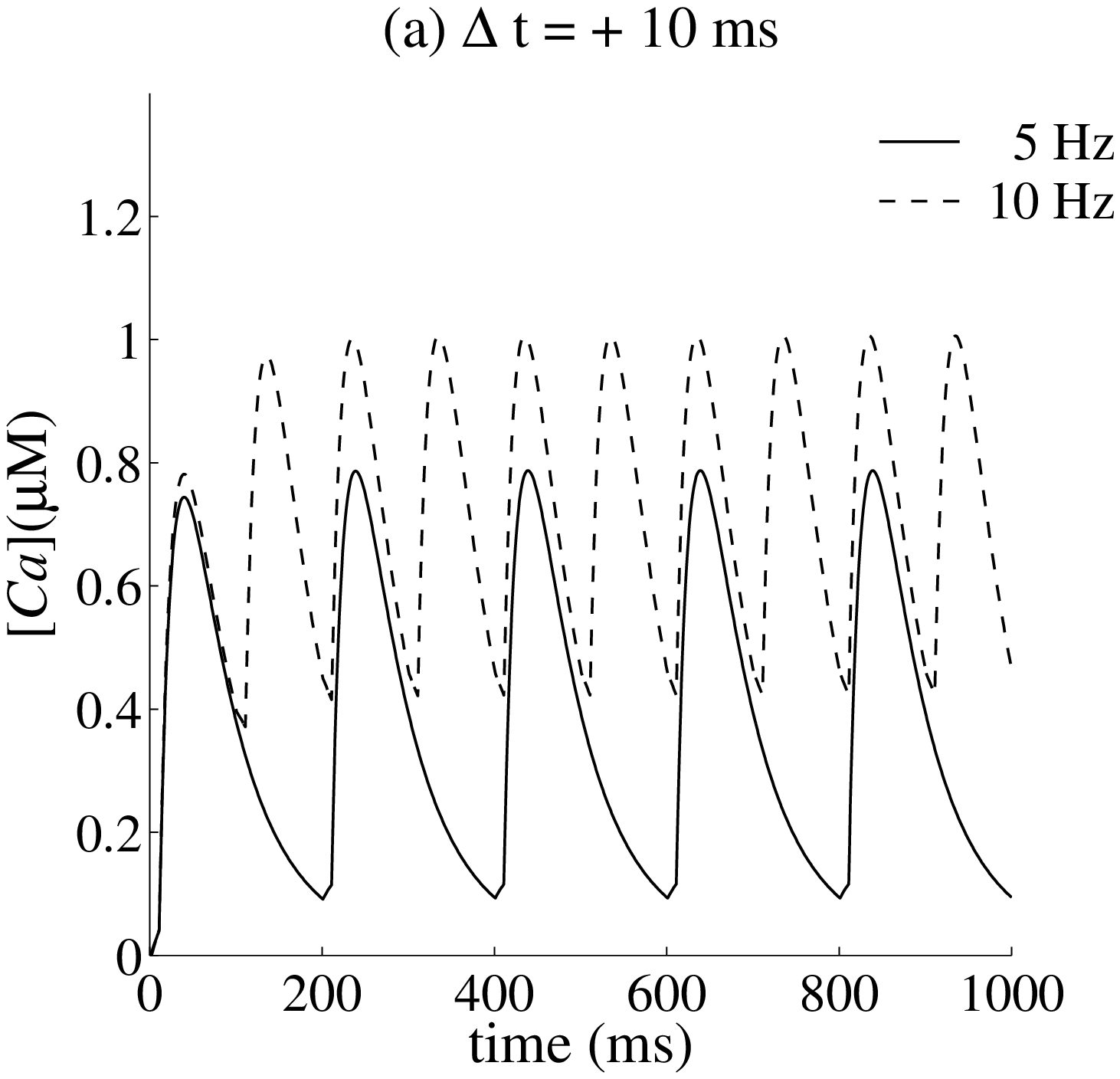}
\hspace{0.5cm}
\includegraphics[width=7cm]{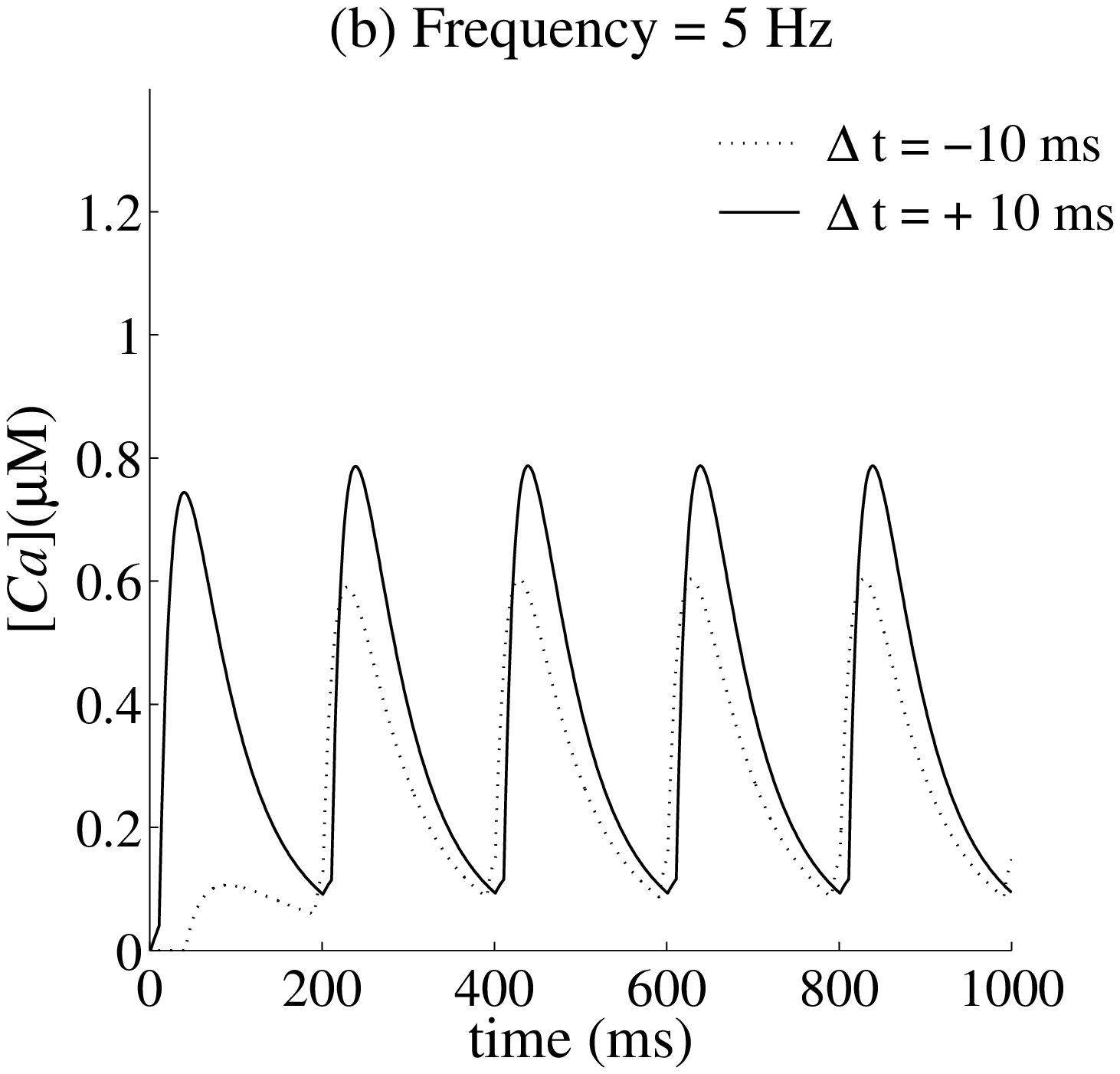}
\caption{\label{fig:rate}
a) Rate-dependence of the calcium transients as derived in expressions \ref{eq:termd} through \ref{eq:termg} for $\Delta t = 10$ ms at 5 Hz (solid line) and 10 Hz (dashed line). b) Calcium transients at 5 Hz for $\Delta t = +10$ ms (solid line) and $\Delta t = -10$ ms (dotted line). All parameters as in FIG. \ref{fig:Ipeak}.
}
\end{figure*}

Plots of these expressions are shown in FIG. \ref{fig:rate}. As expected, the overall calcium is frequency and timing-dependent. There is temporal integration of calcium levels, so that higher frequency stimulations build up the intracellular calcium concentration (FIG. \ref{fig:rate}a). In addition, the $\Delta t$-dependence shown in previous sections (FIG. \ref{fig:rate}b) for single pairs of pre and post spikes is retained. The amount of temporal integration will naturally depend on the specific values of $\tau$, $\tn$ and $\tB$, and so will the difference between the pre-post and the post-pre conditions, as analyzed before. For example, slower dynamics would result in more time integration at moderate frequencies. However, given these constants, it is possible to set a threshold of calcium concentration between LTD and LTP (or between no-plasticity and LTD) that would correspond to low- and high-frequency STDPs, respectively.

We now extract an expression for the dependence of calcium transients on $\Delta t$ and on $\nu$ in the limit $t \rightarrow \infty$. Let $t = t^{pre}_N + \delta$, where the new variable $\delta$ tracks the time since the last presynaptic spike, and define $\gamma_1 = \bar{g}(a+bV_R)\tb [B_{\infty}(1-\mu) + \mu]$ and $\gamma_2 = \bar{g}bV_B\tc [B_\infty (1-\mu)+\mu]$. For $t \rightarrow \infty$ and $\Delta t <\delta$, we have:

\begin{align}
\label{eq:CaDelta1}
[Ca](\delta) = &\gamma_1 e^{-\delta/\tau} \left\{ e^{\delta/\tb} + \left[ \frac{e^{1/\nu \tb} - e^{1/\nu \tau}}{e^{1/\nu \tau} -1 }\right] \right\} \notag \\
& + \gamma_2 e^{-\delta/\tau} e^{\Delta t/\tB} \Bigg\{ \Bigg.\left[ \frac{e^{1/\nu \tc} - e^{\Delta t/\tc}}{e^{1/\nu \tau}-1} \right] \notag \\
& \phantom{-} + \exp \left( \frac{-1}{\nu \tB} + \frac{1}{\nu \tau}\right) \left[ \frac{ e^{\Delta t/\tc} - 1}{e^{1/\nu \tau} - 1} \right] \notag \\
& \phantom{-} + \left( e^{\delta/\tc} - e^{\Delta t/\tc}\right) \Bigg. \Bigg\}
\end{align}

And for $\Delta t > \delta$:
\begin{align}
\label{eq:CaDelta2}
[Ca](\delta) = &\gamma_1 e^{-\delta/\tau} \left\{ e^{\delta/\tb} + \left[ \frac{e^{1/\nu \tb} - e^{1/\nu \tau}}{e^{1/\nu \tau} -1 }\right]\right\} \notag \\
& + \gamma_2 e^{-\delta/\tau} e^{\Delta t/\tB} \Bigg\{ \Bigg.\left[ e^{\Delta t/\tc} - 1 \right] \left[ \frac{e^{-1/\nu \tB}}{e^{1/\nu \tau}-1}\right] \notag \\
& \phantom{-} + \left[ \frac{e^{1/\nu \tc} - e^{\Delta t/\tc}}{e^{1/\nu \tau} - 1} \right] \notag \\
& \phantom{-} + \left( e^{\delta/\tc} - 1 \right) e^{-1/\nu\tB}\Bigg. \Bigg\}
\end{align}

\begin{figure*}
\includegraphics[width=7cm]{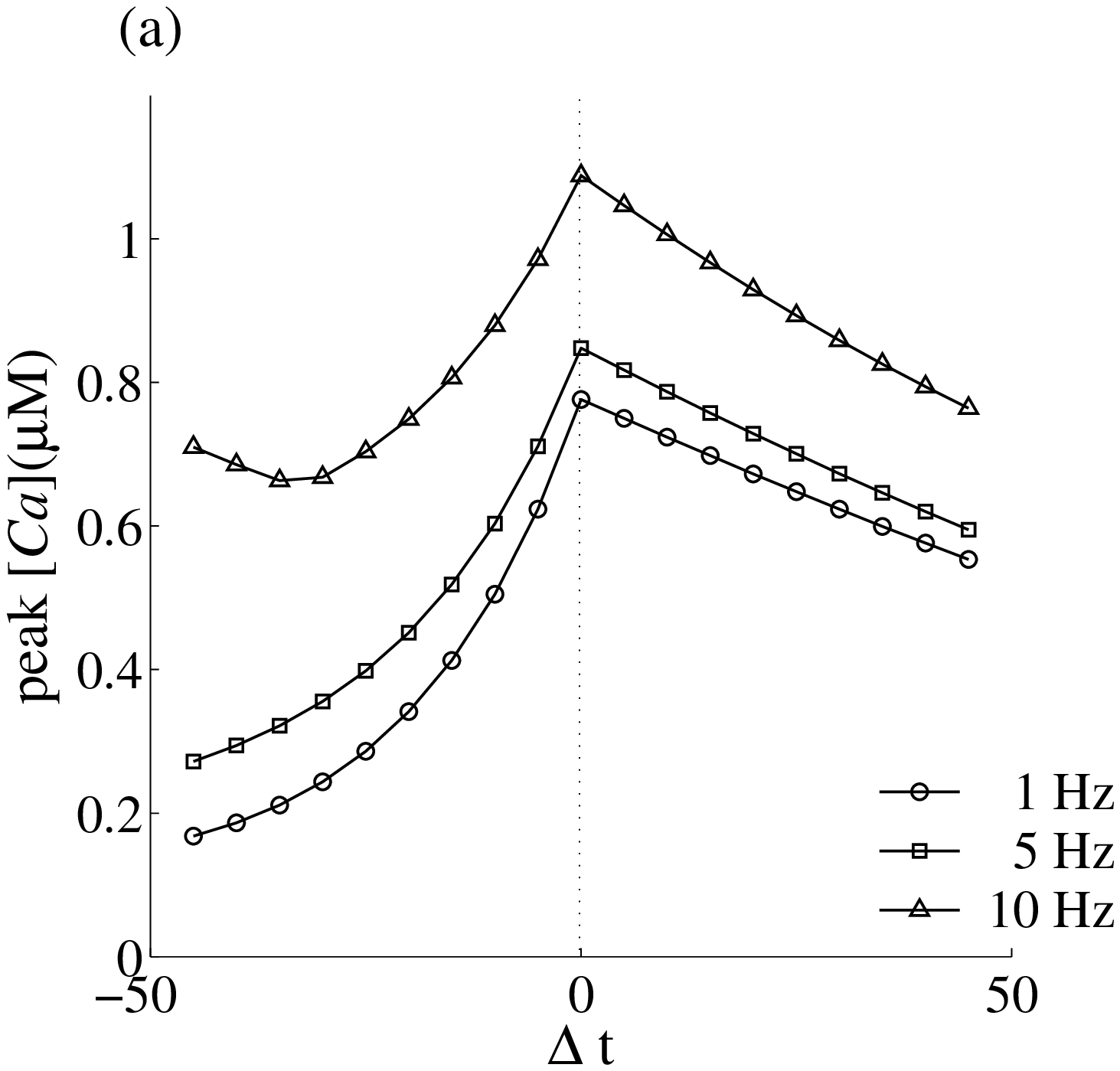}
\hspace{0.5cm}
\includegraphics[width=7cm]{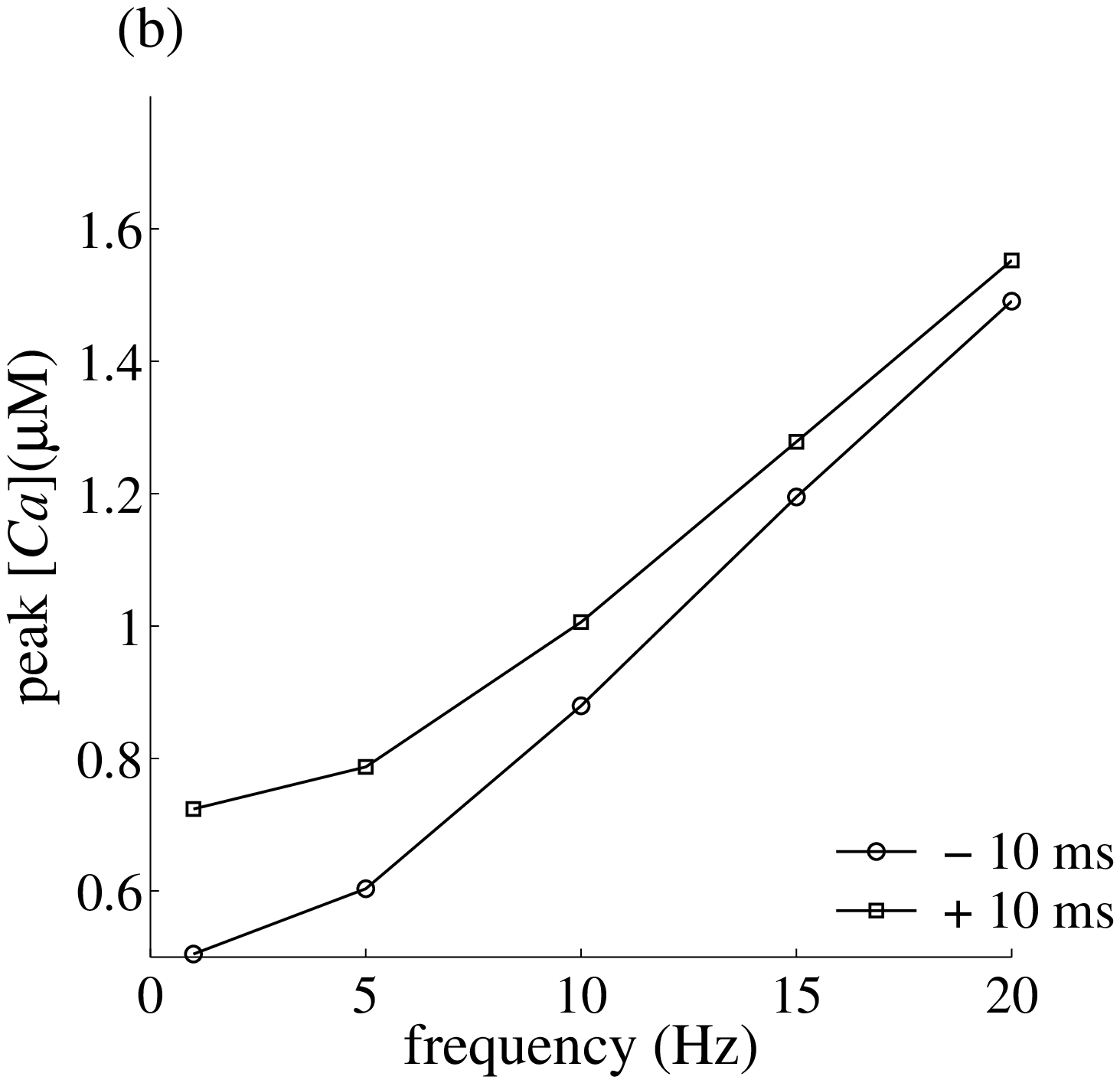}
\caption{\label{fig:PEAK}
a) Peak calcium transients as a function of the timing between pre and postsynaptic spikes for three different frequencies: 1 Hz (circle), 5 Hz (square) and 10 Hz (triangle). b) Peak calcium transients as a function of the frequencies for a pre-post (square) and a post-pre (circle) conditions.
}
\end{figure*}

These expressions show how the peaks of the calcium transients depend on the timing and on the frequency of the pre and postsynaptic spikes. As the frequency increases (FIG. \ref{fig:PEAK}a) the $\Delta t$-dependent curves move up due to temporal integration, and the difference between pre-post and post-pre (FIG. {\ref{fig:PEAK}b}) becomes smaller. 

%%%%%%%%
\section{Stochastic properties of calcium influx \label{sec:var}}
%%%%%%%%

In previous sections we have calculated the mean calcium transients under various conditions and assumptions. However, synaptic transmission is a random process and the trial-by-trial calcium transients deviate from the mean. One source of variability is the stochasticity of the presynaptic neurotransmitter release. A failure of release at a given synapse would result in eliminating the postsynaptic calcium transient at that synapse. Although this source of variability can be significant, its consequences are simple to predict, because the absence of release will result in no synaptic plasticity in the specific spine. Another source of variablity is the stochasticity of the postsynaptic opening of NMDA channels. Because the number of NMDAR at each synapse can be small ($\approx 10$) \cite{RaccaEtAl00}, these fluctuations could be considerable. 

In this section, we calculate the variability due to $\n $ NMDAR in the postsynaptic terminal. We assume the following simple Markov model: the state vector $\pp_n(t)$ denotes the probability of being in each one of the $n$ possible states and the transition matrix $R_n$ represents transition probabilities from state to state. For a Markov processes, the evolution equation has the form:

\begin{equation}
{\frac{d\pp_n}{dt}}=R_n\pp_n,
\end{equation}
\noindent with the solution: 

\begin{equation}
\pp_n(t)=\pp_n(0)\exp\left(\int_{0}^t{R_n dt^\prime}\right),
\end{equation}
\noindent where $\pp_n(0)$ is the initial condition.

In the simplest model the process has only two states, unbound $u$ and open-bound $o$:
\begin{equation}
u
\begin{array}{c}
k_1G\\
\longrightarrow \\
\longleftarrow \\
k_{-1} 
\end{array} o
\end{equation}
\noindent where $G$ is the glutamate concentration, $k_1$ and $k_{-1}$ are the forward and backward time constants respectively.

This model is clearly simplified; typically, a five-state model is used to account for NMDAR kinetics. However, it is sufficient for describing the single exponential decay kinetics we have assumed for the NMDAR since, during decay, $G=0$ (see equation \ref{eq:p_solve}, below). The transition matrix for this model is:

\begin{equation}
R_2=\begin{pmatrix} 
    -k_1G & k_{-1} \\
     k_1G & -k_{-1}.
\end{pmatrix} 
\end{equation}
where $G$ is the neurotransmitter concentration. We assume that $G$ has a constant, non-zero value for only a limited amount of time. We are primarily interested in the falling phase ($G=0$) of the NMDAR current.

We denote by $p^u$ and $p^o$ the probability of being in the unbound and open states respectively. Since $p^u+p^o=1$ the differential equation for $p^o$ reads:

\begin{equation}
\dot{p^o}=k_1G-p^o(k_1G+k_{-1}).
\end{equation}

For a constant value of $G$, the solution of this equation is:
\begin{equation}
p^o(t)={\frac{k_1G}{k_1G+k_{-1}}}+Ce^{-(k_1G+k_{-1})t}.
\label{eq:p_solve}
\end{equation}
where the constant $C$ is determined by the initial conditions.

During the falling phase ($G=0$) the time constant has the value $\tau_\mtn{N}=1/k_{-1}$. The solution takes the form $p^o(t)=\mu e^{-t/\tn}$, where $\mu$ can now be defined more precisely as the open probability at the start of the falling phase. The total calcium current through $\n$ identical NMDAR is $I(t)=\sum_{i=1}^\n s_ig \h(V(t))$, where $g$ is the single channel NMDAR conductance, and $s_i=1$ when receptor in open state 0 when in closed and unbound states. To simplify the notation we can rewrite equation \ref{eq:H} as $\h = V_0+V_1\Theta(t-t^{post})e^{-(t-t^{post})/\tB}$, where $V_0 = a+bV_R$ and $V_1 = b V_B$. 

Calcium concentration is assumed to be governed by equation \ref{eq:Ca}, therefore the average calcium concentration will be:

\begin{equation}
\lav [Ca](t)\rav =ge^{-t/\tau}\int_{-\infty}^t dt^\prime e^{t^\prime/\tau}\lav\sum_{i=1}^\n  s_i(t')\rav \h(t'),
\label{eq:ca_av_one}
\end{equation}
where $\lav ~~ \rav$ denotes averaging over the probability distribution of $s_i(t)$.

However, $\lav \sum s_i(t) \rav = \n \lav s_i(t)\rav =\n p^o(t)$, thus the calcium concentration given $\n$ NMDAR is:

\begin{equation} 
\lav [Ca](t)\rav_\n = \n g e^{-t/\tau}\int_{-\infty}^t dt^\prime e^{t^\prime/\tau}p^o(t^\prime)\h(t^\prime).
\label{eq:ca_av_n}
\end{equation}

Expression \ref{eq:ca_av_n} is similar to equation \ref{eq:Ca_solution} since $I(t)= \n g \h(t) p^o(t)$. We have calculated this integral in section \ref{sec:Ca_dyn} for the different cases (pre, pre-post, post-pre). These assumptions are equivalent to the simple Markov model assumed here. This result is shown in equations \ref{eq:ca_av_one} and \ref{eq:ca_av_n}. We note that $\lav [Ca](t)\rav_\n = \n \lav [Ca](t)\rav_1$.

The variance for $\n$ NMDAR, has the form $\sigma^2_\n(t)=\lav [Ca]^2(t) \rav_\n-\lav [Ca](t) \rav _\n^2$. The second moment of the calcium concentration for $\n$ NMDAR is:
\begin{align}
\lav[Ca]^2(t)\rav_\n = & g^2e^{-2t/\tau}\int_{-\infty}^t dt' dt'' e^{t'/\tau}e^{t''/\tau} \h(t') \h(t'') \notag \\
& \phantom{g^2e^{-2t/\tau}\int} 
\times \lav \sum_{i=1}^\n  s_i(t^\prime)\sum_{j=1}^\n s_j(t^{\prime\prime})\rav  \notag\\
= & g^2 e^{-2t/\tau}\int_{-\infty}^t dt' dt'' e^{t'/\tau}e^{t''/\tau} \h(t')\h(t'') \notag \\
& \phantom{g^2e^{-2t/\tau}\int} 
\times \sum_{i,j=1}^\n p^{oo}_{ij}(t',t''), 
\end{align}
where $p^{oo}_{ij}(t,t^\prime)$ is the joint probability of having channel $i$ open at time $t$ and channel $j$ open at time $t^\prime$. We assume independent identical channels, therefore for $i\ne j$ we have $p^{oo}_{ij}(t,t^\prime)= p^o(t)p^o(t^\prime)$, and when $i=j$, $p^{oo}_{ii}(t,t^\prime)$ is defined as $p^{oo}(t,t^\prime)$. We therefore have that:

\begin{align}
\lav [Ca]^2(t)\rav_\n = & g^2 e^{-2t/\tau}\int_{-\infty}^t dt' dt'' e^{t'/\tau}e^{t''/\tau} \h(t')\h(t'')
  \notag \\
& \times \left[\n(\n-1)p^o(t')p^o(t'') + \n p^{oo}(t',t'')\right] \notag \\
\notag \\
= & g^2 \n(\n-1)\lav [Ca](t)\rav^2 + g^2 \n e^{-2t/\tau} \notag \\ 
& \times \int_{-\infty}^t dt' dt'' e^{t'/\tau}e^{t''/\tau}\h(t')\h(t'')p^{oo}(t', t'')
\end{align}

Thus the variance of calcium concentration for $\n$ NMDAR is:

\begin{align}
\sigma_\n^2(t) = & g^2 \n e^{-2t/\tau} \int_{-\infty}^t dt' dt'' e^{t'/\tau}e^{t''/\tau}\h(t')\h(t'') \notag \\ 
& \phantom{\n e^{-2t/\tau} \int} \times \left[ p^{oo}(t',t'') - p^o(t')p^o(t'')\right] \notag \\
= & g^2 \n e^{-2t/\tau} \int_{-\infty}^t dt' dt'' \h(t')\h(t'') e^{t'/\tau}e^{t''/\tau} p^{oo}(t',t'') \notag \\
& \phantom{\n e^{-2t/\tau} \int} - \n \lav [Ca](t)\rav^2.\label{eq:sigma}
\end{align}

\begin{figure*}
\includegraphics[height=5cm]{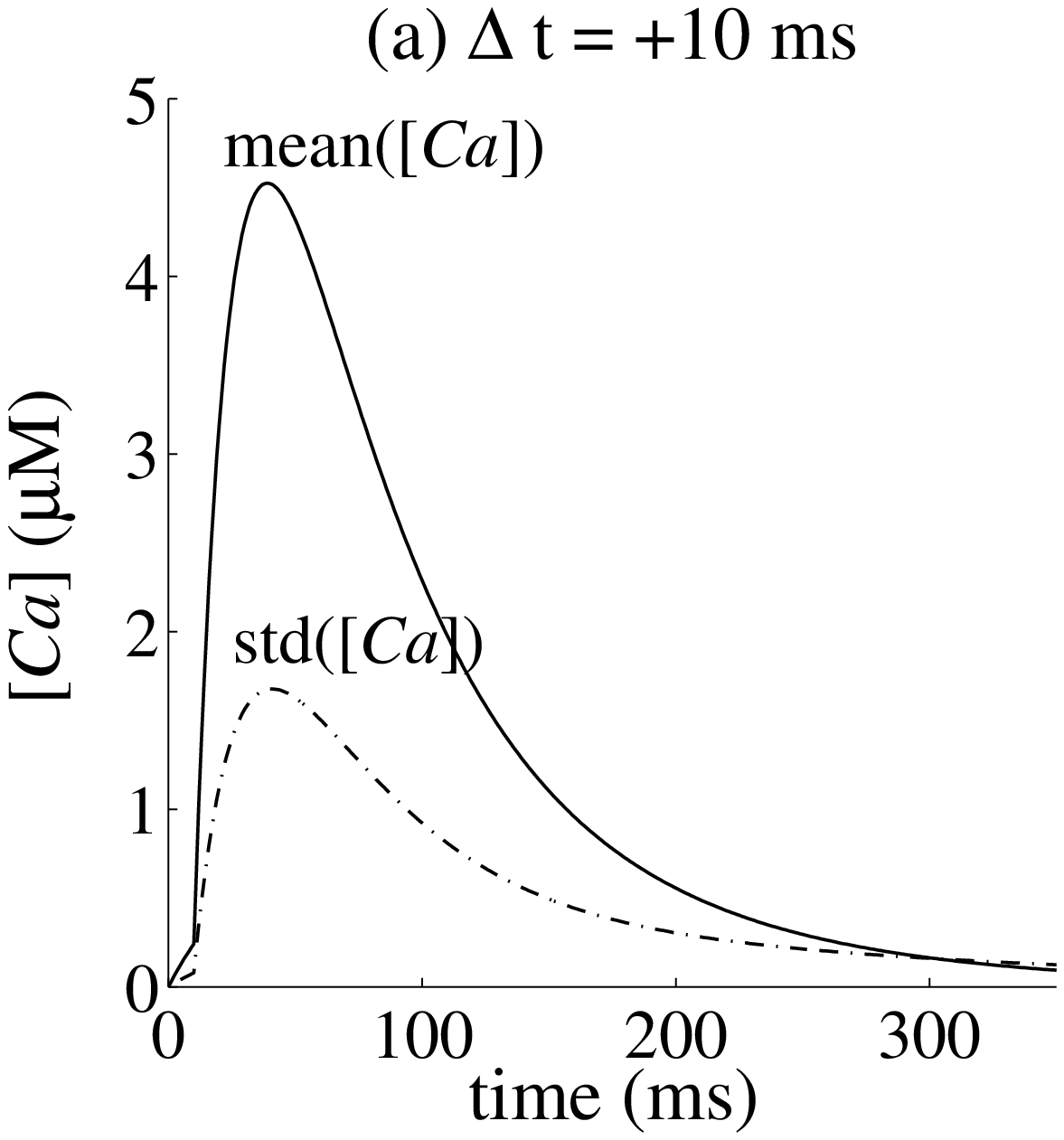}
\includegraphics[height=5cm]{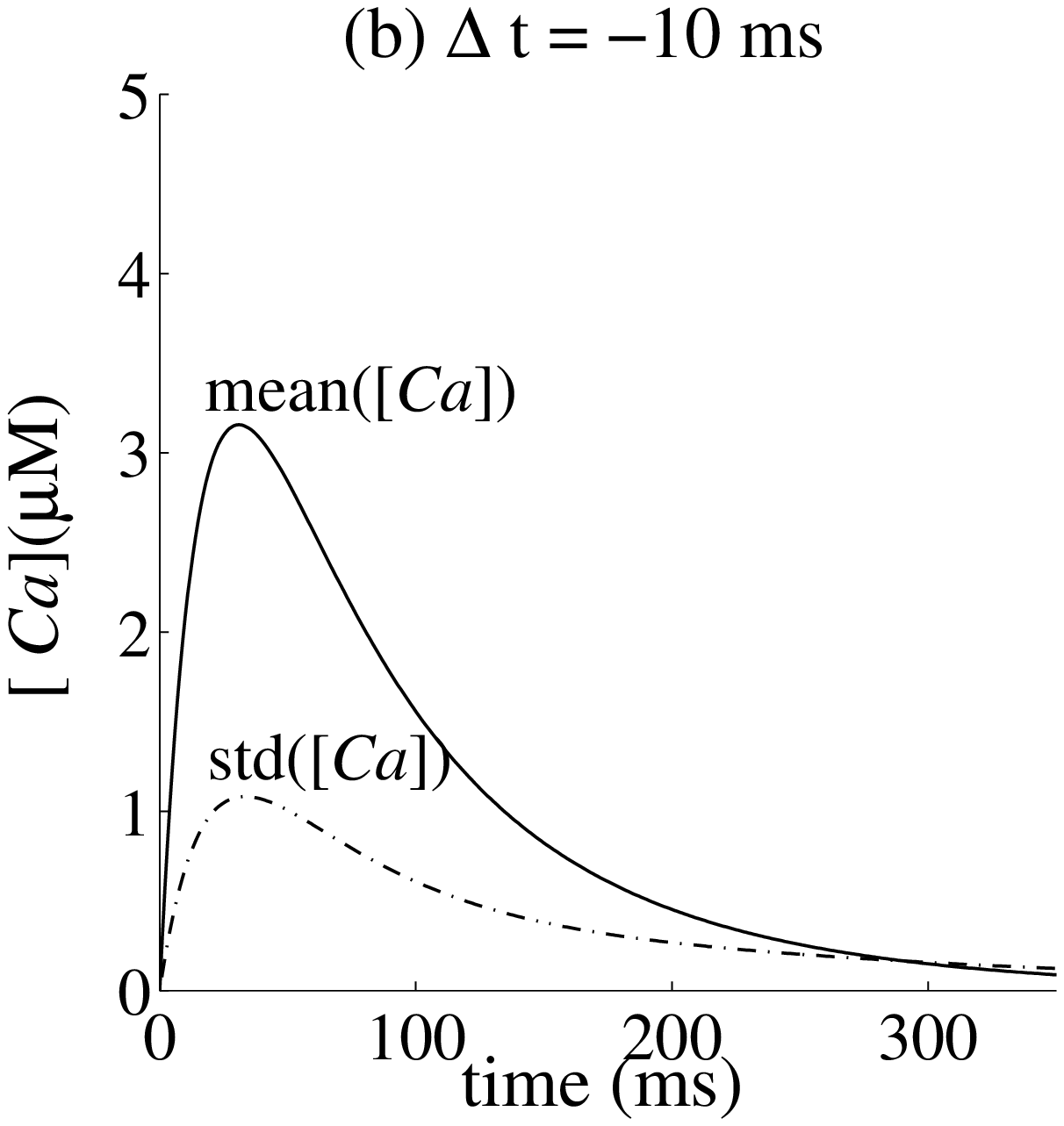}
\includegraphics[height=5cm]{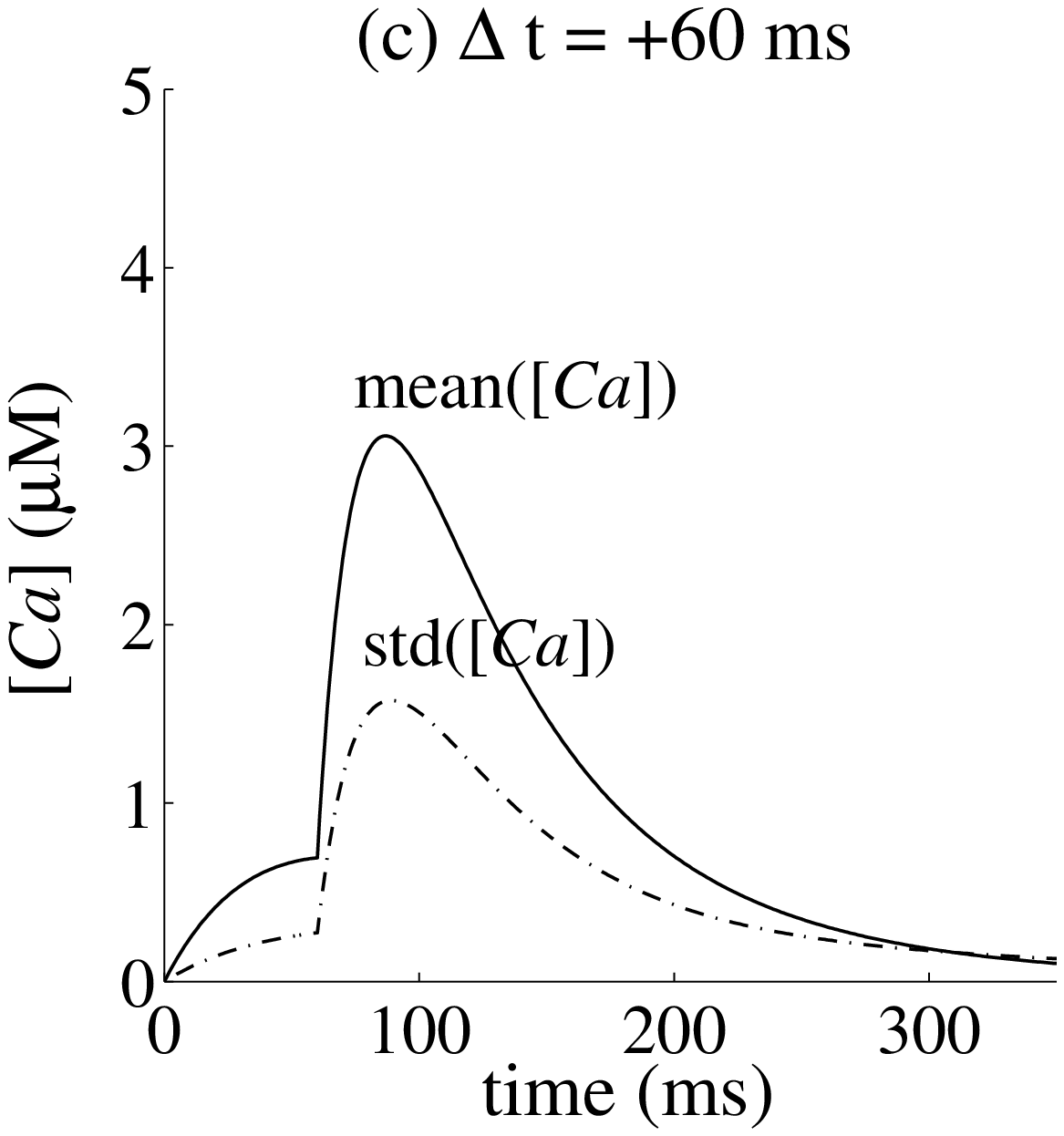}
\caption{
\label{fig:var_calc}
Mean calcium transients (solid line) and their standard deviation (dash-dot line) for different values of $\Delta t$, with $Z=10$ and $\mu = 0.5$. (a) $\Delta t=10$ ms, (b) $\Delta t=-10$ ms, and (c) $\Delta t =60$ ms. The remaining parameters as in FIG. \ref{fig:Ca_dyn}. Note that the peaks of the calcium transients differ from that of FIG. \ref{fig:Ca_dyn} because $\mu$ and $Z$ are different.
}
\end{figure*}

We can use the Bayes rule to rewrite $p^{oo}(t,t')=p^{oo}(t | t')p^o(t')$. Thus there are two types of variables which we need to calculate from the Markov process, $p^o(t)$ and $p^{oo}(t|t^\prime)$. The solution for $p^o(t)$ is given by equation \ref{eq:p_solve}. We define the time $t=0$ as the beginning of the falling phase ($G=0$), or $t^{pre}=0$. Further, we assume an instantaneous rise time. So for $t<0$, $p^o(t)=0$ and at $t=0$ $p^o(0)=\mu$. Thus for $t>0$ $p^o(t)=\mu e^{-t/\tn}$.

Similarly, $p^{oo}(t | t')$ is the probability of being in open state at time $t$ given that the channel was in open state at time $t^\prime$. We calculate this for the case $t,t^\prime\ge 0$. If $t^\prime>t$ this can be solved in a similar manner to $p^o$, since in this regime it is a stationary Markov process, it must be a function of $(t-t')$, and at $t=t$, $p^{oo}(t=t'|t')=1$. During the falling phase there is one directional movement from $o$ to $u$, since $G=0$. Therefore, if we know that at time $t'$ the channel is in an open state it must have been in an open state, at any time $t<t'$. Therefore:

\begin{equation}
p^{oo}(t|t')=\left\{\begin{array}{l@{\quad :\quad}l}e^{-(t-t')/\tn} & t>t'\\
1 & t\le t' \end{array}\right.
\end{equation}

Note that this holds for $(t,t')>0$, and that by definition at times smaller than zero the receptor is unbound. Hence:

\begin{equation}
p^{oo}(t,t^\prime)=\left\{\begin{array}{r@{\quad :\quad}l}\mu\cdot e^{-t/\tau_\mtn{N}} & t>t^\prime \\ \mu \cdot e^{-t^\prime/\tau_\mtn{N}} & t<t^\prime 
\end{array}\right.
\end{equation}

This expression can now be substituted into equation \ref{eq:sigma}, and by using the linear approximation of $\h$ we obtain an expression for  $\sigma_\n^2(t)$ that can be computed analytically. To aid in calculation we denote:

\begin{multline}
\sigma_\n^2(t)=\n\mu g^2e^{-2t/\tau}(V_0^2 \alpha_1 + 2V_0V_1 \alpha_2 + V_1^2 \alpha_3) \\
- \n\lav [Ca](t)\rav^2
\end{multline}
where:

\begin{subequations}
\begin{equation}
\label{eq:A1}
\alpha_1 = \int_0^t dt' \int_0^t dt'' p^{oo}(t',t'')e^{t'/\tau} e^{t''/\tau} \Theta(t)
\end{equation}

\begin{equation}
\alpha_2 = \int_0^t dt' \int_0^t dt'' e^{-(t'-t^{post})/\tB}p^{oo}(t', t'')  
\end{equation}

\begin{align}
\label{eq:A3}
\alpha_3 = & \int_0^t dt' \int_0^t dt'' e^{-(t'-t^{post})/\tn} e^{(t''-t^{post})/\tB} p^{oo}(t', t'') \notag \\
           & \phantom{\int_0^t dt' \int_0^t dt''} \times e^{t'/\tau} e^{t''/\tau} \Theta(t'-t^{post}) \Theta(t''-t^{post}) 
\end{align}
\end{subequations}

In appendix \ref{app:var} we analytically calculate the form of the terms $\alpha_1$, $\alpha_2$ and $\alpha_3$. Although the analytical form is complex, the consequences are simple and significant. As expected, $CV_\n = CV_1/\sqrt{\n}$. Thus the variability is significant only at relatively low $\n$. In FIG. \ref{fig:var_calc} we show $\lav [Ca](t)\rav_\n$ and $\sigma_\n$, for $\Delta t = $ -10, 10 and 60 ms. In all cases both the mean and $\sigma_\n$ change over time. The peak of  $\lav [Ca](t)\rav_\n$ for $\Delta t=-10$ ms and $\Delta t=60$ ms are similar. However, the magnitude of the variability, $\sigma_\n$, is significantly larger at $\Delta t=60$ ms than at $\Delta t=-10$ ms. To quantify the dependence of the relative variability as a function of $\Delta t$, we measure the coefficient of variation, $CV_\n=\sigma_\n/\lav [Ca](t)\rav_\n$, where both $[Ca](t)_\n$ and $\sigma_\n(t)$ are measured at the time $t$ where $\lav [Ca](t)\rav_\n$ is maximal.

\begin{figure}
\includegraphics[width=7cm]{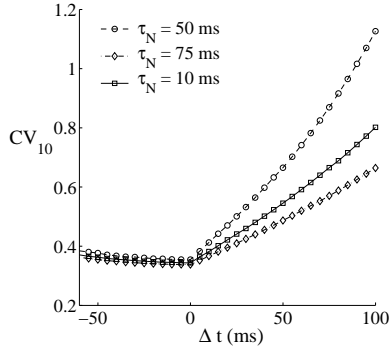}
\caption{
\label{fig:SNR}
Coefficient of variation for $Z = 10$ as a function of $\Delta t$. Notice that the variability decreases with increasing values of $\tn$. Shown are the plots for $\tn$ = 50 ms (circle), $\tn$ = 75 ms (square) and $\tn$ = 100 ms (diamond).
}
\end{figure}

When $\Delta t<0$ the $CV$ is low and decreases as $\Delta t \rightarrow 0$ from below. For $\Delta t>0$ the $CV$ increases as $\Delta t$ increases (FIG \ref{fig:SNR}). It is interesting to compare the $CV$ for values of $\Delta t<0$ and $\Delta t>0$ which have similar peak calcium levels. For example for $\Delta t=-10$ ms, $CV_{10}=0.34$ and for $\Delta t=60$ ms, which has a similar peak calcium level, $CV_{10}=0.51$. Therefore $CV_{10}(60)/CV_{10}(-10)\approx 1.5$. Since $CV_\n=CA_1/\sqrt{\n}$, for this set of parameters, but for every $\n$ $CV_\n(60)/CV_\n(-10)\approx 1.5$. 

\begin{figure}
\includegraphics[width=7cm]{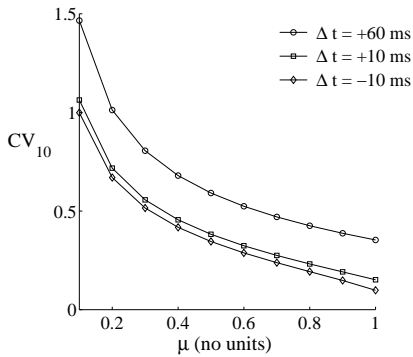}
\caption{
\label{fig:CV_mu}
Coefficient of variation for $Z = 10$ as a function of $\mu$, for three different values of $\Delta t$: $\Delta t=-10$ ms (diamond), $\Delta t=10$ ms (square) and $\Delta t=60$ ms (circle). As $\mu$ increases, CV decreases.
}
\end{figure}

The variability of calcium transients also depends on $\mu$ (FIG. \ref{fig:CV_mu}), decreasing as $\mu$ increases, for all $\Delta t$. The variable $\mu$ is the probability of glutamate binding to postsynaptic receptors given a presynaptic spike, and can be taken to be the presynaptic  probability of release. Thus, as the presynaptic probability of release increases, $CV$ decreases.

%%%%%%%%
\section{Discussion}
%%%%%%%%

Calcium transients due to NMDA currents are believed to play a major role in many forms of synaptic plasticity. We have recently shown that a model based on these transients can account for various different plasticity-induction protocols \cite{ShouvalEtAl02}. In this paper, we compute analytically the calcium dynamics evoked by pairs of pre- and postsynaptic spikes under different conditions, so that the variables that control calcium transients and impact synaptic plasticity can be investigated.

We showed that the peak of the calcium transients depends on the $\Delta t$: for $\Delta t > 0$, the peak concentration decays with the time constant $\tn$ of the NMDAR, and for  $\Delta t < 0$, it decays with the time constant $\tB$ of the BPAP. Therefore, if spike time dependent plasticity, indeed depends on the calcium transients, $\tn$ and $\tB$ will determine the width of the pre-post LTP and the post-pre LTD windows, respectively. These results were confirmed by simulations, in which the approximations we used in the calculations were relaxed. The $\Delta t$ dependence of the peak calcium transient is similar in the more complex and realistic case where the BPAP dynamics is composed of a fast and a slow component: the difference lies in the $\Delta t < 0$ case, which now depends on the combination of the two time constants. The two-component BPAP allows a sharper transition between the post-before-pre and the pre-before-post STDP window, which is further enhanced by a fast calcium dynamics. We also showed that, besides the timing of the pre and post-spikes, the peak values of the calcium transients also depend on the frequency $\nu$ of the pre and postsynaptic conditioning, increasing for greater values of $\nu$. However, at higher frequencies, the dependence on $\Delta t$ decreases. The amount of calcium build-up at a given frequency depends on the system parameters; slower time constants results in more temporal integration. Temporal integration of calcium transients, as described here, can explain why, in some cases, the induction of STDP is frequency-dependent.

One of the predictions of the unified calcium model is that there exists a pre-before-post LTD for values of $\Delta t$ greater than the ones that elicit LTP \cite{ShouvalEtAl02}. This has been shown experimentally by some investigators \cite{NishiyamaEtAl00}; others have only placed a small number of data points in this region \cite{BiPoo98,SojstromEtAl01}. In all published STDP experiments, there is a large variability in the magnitude and the sign of synaptic plasticity across the different values of $\Delta t$. It would therefore be difficult to assess the existence of this type of LTD without a large amount of data. The variability encountered in the experiments indicates that it should be important to examine the stochastic properties of the calcium transients, in addition to their mean.

We calculated the variance of the calcium transients and showed that for a small number $Z$ of NMDAR, it can be quite significant. There are indications that the number of NMDAR in the dendrite is indeed small ($\approx 10$) \cite{RaccaEtAl00}. Further, we showed that the $CV$ increases monotonically with $\Delta t$, for $\Delta t>0$. These fluctuations can have significant implications on the outcome of downstream processes dependent on calcium, such as phosphorylation or other second-messenger cascades upon which plasticity relies. 

Previously, several authors have simulated and analyzed dynamical models of synaptic plasticity that can lead to STDP \cite{KempterEtAl99,SennEtAl01,SongEtAl2000,LevyEtAl01,AbarbanelEtAl02}. 
These models typically bypass detailed descriptions of the physiology and biochemistry underlying synaptic plasticity. Recently, biophysical models have been receiving increasing focus \cite{CastellaniEtAl01,KitajimaHara00,ShouvalEtAl02,KarmarkarBuonomano02,ShouvalEtAl02b}. These more mechanistic approaches have allowed one to show that various different induction protocols can be accounted for by a single set of assumptions regarding the dependence of plasticity on calcium concentrations \cite{ShouvalEtAl02}. In these models, plasticity is a highly non-linear function of calcium. Therefore, fluctuations of calcium transients will not only add spread to the resulting LTD and LTP, but will also shift their mean (the mean of the solution is not equivalent to the solution given the mean calcium level). For example, in the unified calcium model \cite{ShouvalEtAl02}, the sign and magnitude of synaptic plasticity is determined by a saturated U-shaped function of calcium concentration. When calcium falls below a threshold $\theta_d$, no plasticity occurs, when it falls between $\theta_d$ and $\theta_p$, $\theta_p > \theta_d$, LTD is induced, and for calcium above $\theta_p$, LTP is induced. Thus, in the LTD region, if the spread of the calcium is of the order of $\theta_p - \theta_d$, occasional LTP can still happen. We showed an explicit example for the cases of $\Delta t = -10$ ms and $\Delta t = 60$ ms, where the mean peak calcium concentrations are similar, but the variability for the latter is larger. If this mean calcium is in the LTD range, it is likely that the observed phenomenon is that the LTD is stronger and more robust for $\Delta t = -10$ ms than for $\Delta t = 60$ ms.

This work is based on several simplifying assumptions. This allows us a qualitative understanding of the aspects of calcium dynamics that are crucial for the induction of synaptic plasticity. Incorporation of more realistic features, such as those assumed in the unified calcium model, would probably yield quantitatively more precise results. For example, a double-component NMDAR decay would allow for more calcium build-up without sacrificing the temporal asymmetry between pre-post and post-pre conditions. However, we believe that, despite the constraints of analytical tractability, this study can serve as a basis for future models that aim to formalize the biophysical basis of synaptic plasticity.

%%%%%%%%
\section{Acknowledgments}
%%%%%%%%

HZS would like to thank the members of the IBNS at Brown University as well as M. Tsodyks, D. Golomb and Y. Kubota for useful discussions and comments, and the ITP at Santa Barbara where some of this work was carried out. This work was partly supported by the Burroughs-Wellcome Fund and the Charles A. Dana Foundation.

%%%%%%%%%

\appendix

%%%%%%%%%

\section{Variance calculation \label{app:var}}

Using the definition in equation \ref{eq:A1} through \ref{eq:A3} we obtain:

\begin{align}
\alpha_1 = & \left[\int_0^t dt' \int_{t'}^t dt'' e^{-t''/\tn} e^{t'/\tau} e^{t''/\tau} \right. \notag \\
           & \left. + \int_0^t dt' \int_0^{t'} dt'' e^{-t'/\tn} e^{t'/\tau} e^{t''/\tau}\right] \Theta(t) \notag \\
         = & \left[\int_0^t dt' e^{t'/\tau} \tb \left(e^{t/\tb}-e^{t'/\tb}\right) \right. \notag \\
           & \left. + \int_0^t dt' e^{t'/\tb} \tau \left(e^{t'/\tau}-1 \right) dt' \right] \Theta(t) \notag \\
         = & \left[ e^{t/\tb} \tb \tau \left((e^{t/\tau} - 1 \right)- \tb \int_0^t dt' e^{t'/\td}\right. \notag \\
           & + \left. \tau \int_0^t dt' e^{t^\prime/\td}-\tau \int_0^t dt' e^{t'/\tb} \right] \Theta(t) \notag\\
         = & \left[ e^{t/\td} \left(\tb\tau - \tb\td + \tau \td \right) \right. \notag \\
           & \left. -2 e^{t/\tb} \tb\tau + \left(\tb\td - \tau\td + \tb\tau \right) \right] \Theta(t),
\end{align}
where $\td^{-1} = 2 \tau^{-1} - \tn^{-1}$. 
\begin{align}
\alpha_2 = & e^{t^{post}/\tB}\left[ \te(\tau-\tb)(e^{t/\te}-e^{\tilde{t}/\te}) - \right. \notag\\
           & \left. - \tau \tc (e^{t/\tc}-e^{\tilde{t}/\tc})+ \tb\tf e^{t/\tb}(e^{t/\tf}-e^{\tilde{t}/\tf})\right] \notag\\
           & \times \Theta(t-\tilde{t})
\end{align}
where $\tc$ is as defined before, $\te^{-1} = 2\tau^{-1} - \tn^{-1} - \tB^{-1}$, $\tf^{-1} = \tau^{-1} - \tB^{-1}$ and

\begin{align}
\tilde{t}=\left\{\begin{array}{r@{\quad :\quad}l}t^{post} & t^{post}>0 \\ 0 & t^{post} \le 0 
\end{array}\right.
\end{align}

Finally,
\begin{align} 
\alpha_3 = 2 e^{2 t^{post}/\tB} \tf \tc &\left(e^{t/\tf} - e^{t/\tc} e^{\tilde{t}/\tf} + e^{\tilde{t}/\tf}\right) \notag \\
&\times \Theta(t-\tilde{t})
\end{align}

\begin{figure}
\includegraphics[width=8.5cm]{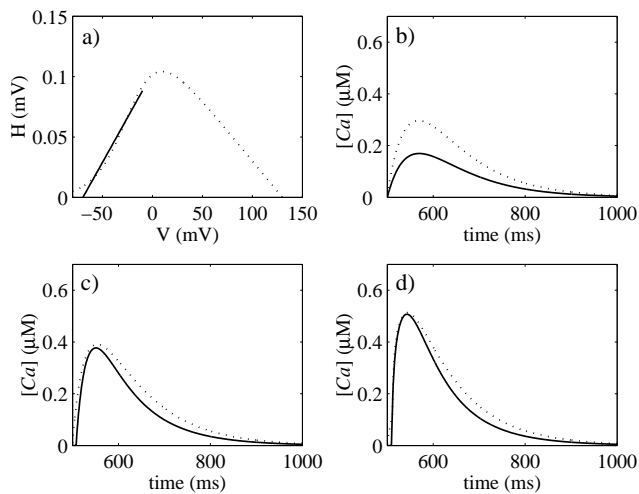}
\caption{\label{fig:comparison}
The error estimation for the linear approximation of the $\h$-function. a) The original $\h$-function (dotted line) and its linear approximation (solid line) in the interval $[-70, -10]$ mV. b), c) and d) The calcium transients elicited by the pre-only, post-pre and pre-post conditions, respectively, for the full expression (dotted line) and for the linear approximation (solid line).
}

\end{figure}

%%%%%%%%
\section{Numerical estimation of the error resulting from the linear approximation of $\h$ \label{app:B}}
%%%%%%%%

In the simulations, we used the following form for the voltage-dependence of the NMDAR.:

\begin{equation}
\h(V) = \frac{V-V_{rev}}{1 + \frac{e^{-0.062 V}}{3.57}}
\label{eq:hV}
\end{equation}
\noindent where $ V_{rev} = 130$ mV, fit performed in the interval $[-70, -10]$ mV.

Such functional dependence captures the qualitative non-linear dependence of the NMDAR on the voltage, although the reversal potential for calcium channels might be inaccurate at the values close to it. In our analysis, we approximated $\h$ with a linear dependence on voltage. The fit used is shown in FIG. \ref{fig:comparison}a. We see that over the relevant range ([-70 0]) this is a relatively good approximation.

In panels b-d of FIG. \ref{fig:comparison} we compare the numerically extracted calcium transients (dashed line), using the non-linear $\h$ and the analytical results obtained for the linear approximation (solid line). 
At rest (FIG. \ref{fig:comparison}b), the linear approximation significantly underestimates the magnitudes of the transients, because this is where the linear fit departs the most from the true curve. However, in the working regime ([-10, 10] mV), the approximation can be considered adequate. 

Let equation \ref{eq:hV} be rewritten as $\h (v) = a + bV$ as in equation \ref{eq:H}. The parameters of linearization are $a =  0.1031/\bar{g}$ and $b = 0.0015/\bar{g}$, where $\bar{g} = -10^{-3}$ is the average NMDA conductance.

\end{document}